\documentclass{aa}  
\usepackage{graphicx}
\usepackage{txfonts}

\newcommand{\PaperI}{\href{https://doi.org/10.1051/0004-6361/202558739}{Paper~I}}

\newcommand{\pc}{\>{\mathrm{pc}}}
\newcommand{\kpc}{\mbox{$\>{\mathrm{kpc}}$}} 

\newcommand{\Gyr}{\mbox{$\>{\mathrm{Gyr}}$}}

\newcommand\degree{^\circ}

\newcommand{\feh}{\mbox{$\mathrm{[Fe/H]}$}}

\newcommand\lv{{$l_{\mathrm{v}}$}}
\mathchardef\mhyphen="2D

\usepackage{natbib,twoopt}
\usepackage[colorlinks,breaklinks=true]{hyperref}
\hypersetup{
    citecolor={blue},
    linkcolor={blue},
    urlcolor={blue}}
\bibpunct{(}{)}{;}{a}{}{,}             
\makeatletter
  \newcommandtwoopt{\citeads}[3][][]{\href{http://adsabs.harvard.edu/abs/#3}%
    {\def\hyper@linkstart##1##2{}%
     \let\hyper@linkend\@empty\citealp[#1][#2]{#3}}}
  \newcommandtwoopt{\citepads}[3][][]{\href{http://adsabs.harvard.edu/abs/#3}%
    {\def\hyper@linkstart##1##2{}%
     \let\hyper@linkend\@empty\citep[#1][#2]{#3}}}
  \newcommandtwoopt{\citetads}[3][][]{\href{http://adsabs.harvard.edu/abs/#3}%
    {\def\hyper@linkstart##1##2{}%
     \let\hyper@linkend\@empty\citet[#1][#2]{#3}}}
  \newcommandtwoopt{\citeyearads}[3][][]%
    {\href{http://adsabs.harvard.edu/abs/#3}
    {\def\hyper@linkstart##1##2{}%
     \let\hyper@linkend\@empty\citeyear[#1][#2]{#3}}}
\makeatother

\usepackage[dvipsnames]{xcolor}

\definecolor{dgreen}{RGB}{0,150,40}

\usepackage{soul}

\usepackage[normalem]{ulem}
\usepackage{xcolor}

\begin{document}

   \title{Kinematic hints of a nuclear bar in the Milky Way}

   \author{K. Fiteni
          \inst{1,2}
          \and
          M.C. Sormani\inst{1}
          \and
           V. P. Debattista\inst{3}
           \and
          F. Nogueras-Lara\inst{4}
          \and
          R. Schödel\inst{4}
          \and
          J. L. Sanders\inst{5}
          \and
          M. Schultheis\inst{6}
          \and
          X. Li\inst{1}
          \and
          A. Vasini\inst{1}
          \and
          Z.-X. Feng\inst{1}
          \and
          M. Donati\inst{1}
          }

   \institute{Como Lake centre for AstroPhysics (CLAP), DiSAT, Università dell’Insubria, Via Valleggio 11, 22100 Como, Italy\\
   \email{karlfiteni@gmail.com}
   \and
     Institute of Space Sciences \& Astronomy, University of Malta, Msida MSD 2080, Malta
   \and
     Jeremiah Horrocks Institute, University of Lancashire, Preston PR1 2HE, UK
   \and
     Instituto de Astrof\'isica de Andaluc\'ia (CSIC), Glorieta de la Astronom\'ia s/n, E-18008 Granada, Spain
   \and
    Department of Physics and Astronomy, University College London, London WC1E 6BT, UK
    \and
    Universit\'e C\^ote d’Azur, Observatoire de la C\^ote d’Azur, Laboratoire Lagrange, CNRS, Blvd de l’Observatoire, 06304 Nice, France
}

   \date{Received January 27, 2026; accepted xxx xx, 202x}

  \abstract
   {The Milky Way hosts a flattened nuclear stellar disc (NSD) that dominates the gravitational potential in the inner few hundred parsecs, but whether the NSD is purely axisymmetric or contains a nuclear bar remains an open question. We test for the presence of a nuclear bar with kinematic diagnostics, combining positions and line-of-sight velocities from the KMOS NSD survey with proper motions from VIRAC2 to construct the $(v_\ell, v_\mathrm{los})$ velocity ellipse. We measure the vertex deviation $l_v$ and the anisotropy $\beta$ of the velocity ellipse for a number of sub-samples after applying quality cuts designed to minimise contamination from large-scale bar stars. For our primary sample, comprising stars within $|\ell| < 0.9^\circ$, $-0.4^\circ < b < 0.25^\circ$, $\mathrm{[Fe/H]} > -0.3$, we measure a negative vertex deviation $l_v = -54.8^{+13.1}_{-14.8}\,^\circ$ with moderate anisotropy $\beta = 0.16^{+0.08}_{-0.05}$. A subsample restricted to the innermost four fields yields $l_v = -64.3^{+12.1}_{-12.2}\,^\circ$ with higher anisotropy $\beta = 0.38^{+0.12}_{-0.07}$. The direction of maximum velocity dispersion is oriented along Galactic longitude, opposite to what is observed in samples dominated by large-scale bar stars, where line-of-sight dispersion dominates. We verify the robustness of these signatures against extinction-driven photometric incompleteness, primary-bar contamination, and the choice of metallicity threshold used to suppress bar interlopers; the kinematic measurements remain stable in all cases. The signatures are inconsistent with a NSD which is axisymmetric or oriented orthogonally to the primary bar, but match expectations for a nuclear bar oriented at $\alpha \approx 60^\circ$--$75^\circ$ to the Sun--Galactic-Centre line, with its near side pointing toward positive Galactic longitude. While definitive confirmation awaits the larger and more precise samples expected from upcoming surveys, our results provide the first kinematic indication of a possible nuclear bar in the Milky Way.}


   \maketitle
%
\section{Introduction}\label{intro}

Bars are a prevalent structural feature in disc galaxies, with approximately $60\%$ of disc galaxies hosting a large-scale bar \citep{Eskridge+2000}. The Milky Way (MW) itself hosts a prominent large-scale bar, whose major axis is oriented at an angle of approximately $27\degree$ relative to the Sun-Galactic Centre line, and a semi-major axis $\approx 5 \kpc$ \citep[see][and references therein]{Bland-Hawthorn+2016}. Early studies of external galaxies based on photographic plates also identified the existence of double-barred systems, where a barred galaxy contains a secondary, nuclear bar \citep{Vaucouleurs+1975, Sandage+1979}. Subsequent observational studies sought to quantify the frequency and properties of nuclear bars, finding that roughly $20\%$ of barred galaxies host a nuclear bar \citep[e.g.,][]{Laine+2002, Erwin+2011, Erwin+2024}. In addition, approximately $50\%$ of barred galaxies with MW masses and above ($\log(M_{\star}/M_{\odot})>10.5$) are double-barred, with nuclear bar sizes typically a few hundred parsecs \citep{Erwin+2024}. Studies also find that inner and outer bars are dynamically decoupled, having orientations and pattern speeds independent of each other \citep{Buta+1993, Friedli+1993, Maciejewski+2000, Corsini+2003, Debattista+2007, Du+2015, Li+2023}.

There have been previous claims that the MW is double-barred. \citet{Alard+2001} used H- and K-band star counts from 2MASS to derive the projected stellar density in the bulge region. They reported an overdensity at negative longitudes which they interpreted as a signature of a nuclear bar. \citet{Rodriguez+2008} derived the mass distribution of the inner Galaxy by fitting the 2MASS star count map to a model including a disc, bulge and nuclear bar. They then simulated the gas dynamics in the deduced gravitational potential, recovering an orientation of the nuclear bar of $\alpha \approx 60\degree \mhyphen 75\degree$, where $\alpha$ is measured from the Sun-GC line, with positive $\alpha$ in the positive longitude direction. \cite{Nishiyama+2005} used red clump (RC) stars at a latitude of $b=+1\degree$ to trace the main bar orientation as a function of longitude ($-10.5<\ell\ [\degree]<10.5$). They reported a change in the inclination angle for the Galactic bar at fields with $|l| < 4\degree$, which they interpreted as the effect of a distinct nuclear bar. \citet{Gonzalez+2011} later confirmed this result using RCs from the VISTA Variables in the Via Lactea (VVV) survey to trace the orientation of the bar at latitudes $b = \pm 1\degree$. While the works above appear to provide compelling evidence of a nuclear bar, \citet{Gerhard+2012} used N-body simulations to demonstrate that both the asymmetry in the star count maps and the change in the bar orientation can be reproduced by the radially varying axis ratio of the primary bar even in the absence of a nuclear bar, casting doubts on claims of a nuclear bar based on star counts.

A significant challenge in assessing whether the MW is double-barred stems from our edge-on perspective of the Galactic centre, combined with the substantial, non-uniform extinction throughout the nuclear region \citep[e.g.][]{Bally+1988, Gonzalez+2012, Schultheis+2014,Nogueras-lara+2021,Henshaw+2023}. In particular, roughly 2/3 of the dust/gas near the mid-plane within $|\ell|<1.5\degree$ resides at positive longitudes \citep{Bally+1988}. This complicates the use of star counts and photometric methods by producing an apparent deficit of stars at positive longitudes. If uncorrected, this bias could be misinterpreted as evidence for a nuclear bar oriented with its near side at negative longitudes, a configuration similar to that proposed by \citet{Alard+2001}. 
\begin{figure}
   \centering
    \includegraphics[width=\linewidth]{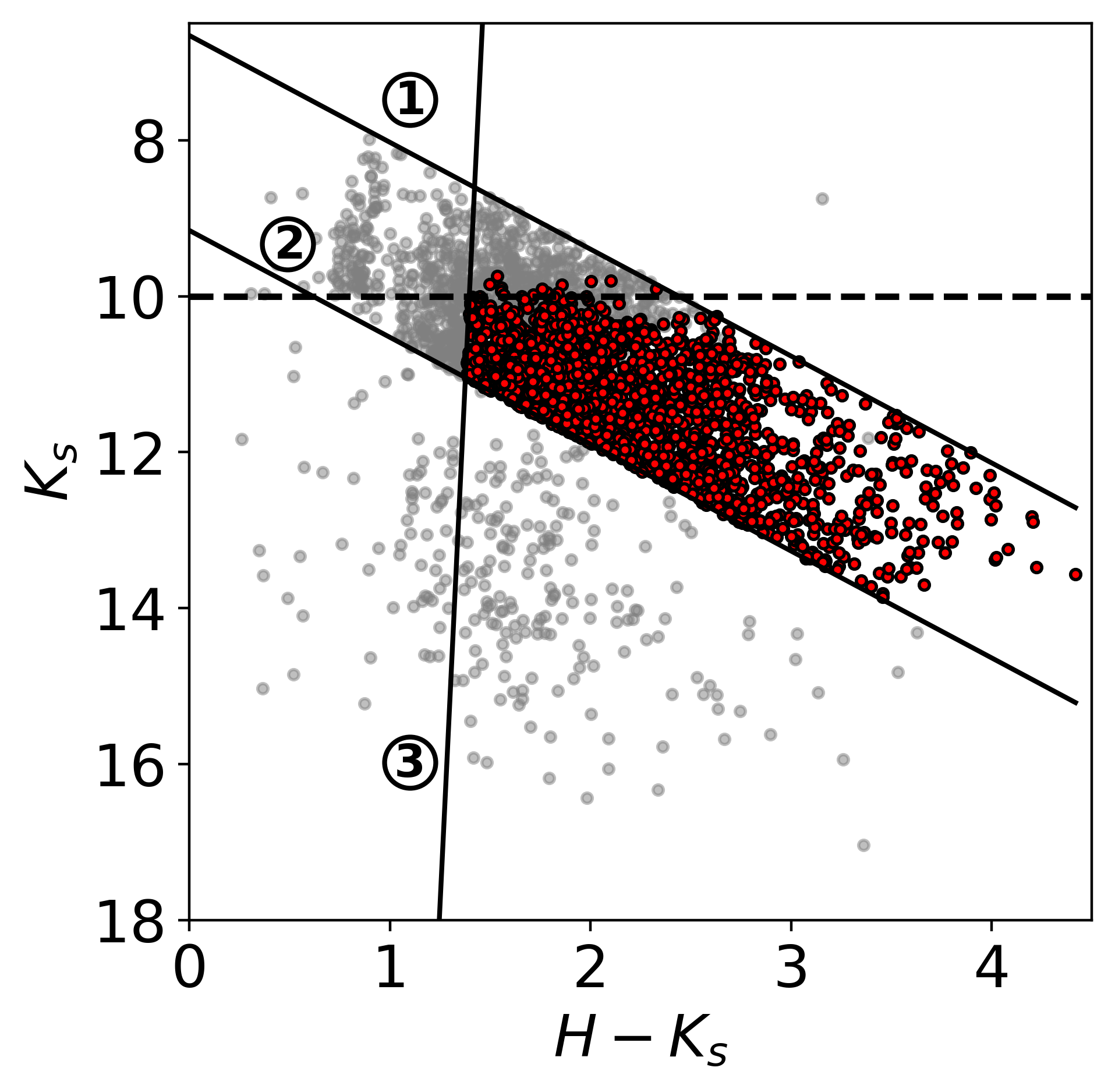}
    \caption{Colour--magnitude diagram $K_s$ versus $H-K_s$ for the full KMOS+VIRAC2 sample. Line 3 reflects the cut $(H-K_s) > \max(1.3,\,-0.0233\,K_s + 1.63)$ that reduces contamination from foreground stars in the sample. Lines 1 and 2 encompassing $6.6575 < K_s - 1.37(H-K_s) < 9.1575$ represent the \citet{Fritz+2021} selection of targets. The horizontal dashed line at $K_s = 10$~mag marks the saturation limit of the VIRAC2 photometric catalogue: stars brighter than this are excluded by the catalogue and consequently absent from our sample (see Sec.~\ref{sec:ext_vertex}).}
    \label{fig:selection}
\end{figure}
An alternative approach to detect a nuclear bar in the MW is to use stellar kinematics. 
Using N-body simulations, \citet{Fiteni+2026}  (henceforth referred to as \PaperI) carried out a systematic investigation into the potential kinematic signatures of a nuclear stellar bar in the MW. They proposed two primary diagnostics: (1) the vertex deviation, $l_v$, of the ($v_{\ell},v_{\mathrm{los}})$ velocity ellipsoid, and (2) the asymmetry in the $\mu_{\ell}$ vs $\ell$ distribution. Their analysis indicated that out of the two diagnostics, the vertex deviation is the most robust, owing to its relative insensitivity to extinction, and the ability to combine stars from different fields across $\ell-b$ space, which can improve the statistical significance of the \lv\ measurement. In particular, they find that $N \gtrsim 500$ stars when combining the inner ($|\ell|<0.9\degree,-0.4\degree<b<0.25\degree$) fields in the KMOS survey fields \citep{Fritz+2021} could be sufficient to reliably compute \lv. In this paper, we apply the vertex-deviation diagnostic established in \PaperI\ to the currently available KMOS+VIRAC2 observations of the NSD region, with the aim of testing whether the existing data already show kinematic evidence of a nuclear bar in the MW.

This paper is organised as follows. In Sec.~\ref{sec:data}, we present the dataset used in our analysis. In Sec.~\ref{sec:vertex_dev} we give a brief outline of the kinematic parameters computed in the study. We then present measurements of the vertex deviation at the Galactic centre in Sec.~\ref{sec:results}. In Sec.~\ref{sec:monte_carlo} we assess the impact of distance and velocity uncertainties on the results. Sections~\ref{sec:extinction} and~\ref{sec:contamination} test the robustness of the kinematic measurement against extinction-driven incompleteness and primary-bar contamination, respectively. We discuss the implications and conclude in Sec.~\ref{sec:summary}.

\section{The observational data}\label{sec:data}

We employ the same dataset used by \citet{Sormani+2022}, which includes line-of-sight velocity measurements from the KMOS K-band spectroscopic survey of the NSD \citep{Fritz+2021}, supplemented by proper motions from the VIRAC2 \citep{Smith+2025} astrometric and photometric catalogue, derived from the Vista Variables in the Via Lactea (VVV) survey \citep{Minniti+2010}. Our initial dataset, consisting of stars having $\mu_{\ell}$ and $v_{\mathrm{los}}$, has a total of $2533$ stars.

\begin{figure}
   \centering
    \includegraphics[width=\linewidth]{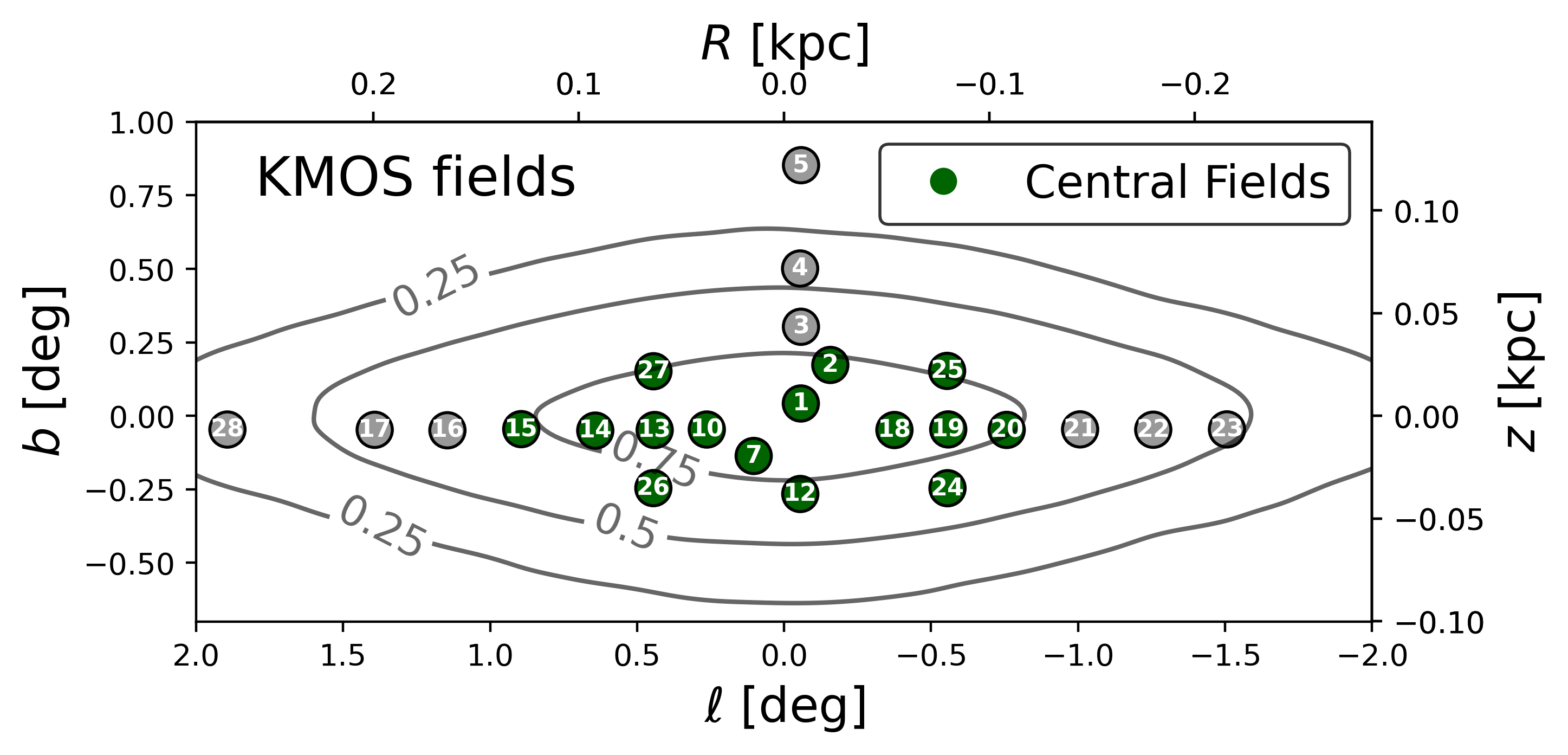}
    \caption{The $(l,b)$ positions of the observational fields in the KMOS NSD survey  of \citet{Fritz+2021}, numbered according to their table A.1. Overlaid are contours reflecting the prominence of the NSD based on the model of the NSD of \citet{Sormani+2022}, with the central (green) fields exhibiting the least ($\lesssim 25\%$) contamination from primary bar stars.}
    \label{fig:kmos_fields}
\end{figure}

Fig.~\ref{fig:selection} shows the Color-Magnitude diagram (CMD) for the raw sample (gray markers). The survey of \citet{Fritz+2021} selects only stars in the region $6.6575 < K_s-1.37 \times (H-K_s) < 9.1575$ (i.e. between lines 1 and 2 in Fig.~\ref{fig:selection}). Foreground stars along the line of sight, originating from both the spiral arms and the Galactic bulge, are subject to substantially different extinction compared to stars located in the central regions \citep[e.g.,][]{Nogueras-lara+2021}. Following \citet{Schultheis+2021}, \cite{Nogueras+2024}, and \citet{Nieuwunster+2024}, we exclude stars with $(H-K_s) > \mathrm{max}(1.3, -0.0233 K_s+1.63)$ (line 3 in Fig.~\ref{fig:selection}). This effectively removes stars in the bulge field (field 5 in Fig.~\ref{fig:kmos_fields}). All subsequent samples are constructed from stars satisfying these selection criteria (red markers). We also note that the VIRAC2 photometric catalogue saturates at $K_s \approx 10$~mag, truncating the bright end of our photometric sample.

In Fig.~\ref{fig:kmos_fields} we show the $(\ell,b)$ positions of the KMOS fields, numbered according to table A.1 in \citet{Fritz+2021}. The overlaid contours show the ratio of the surface mass density of the NSD to the combined surface mass density of the NSD and the primary bar (\PaperI), defined by:

\begin{equation}
    f_{\mathrm{NSD}}=\frac{\Sigma_{\mathrm{NSD}}}{(\Sigma_{\mathrm{NSD}}+\Sigma_{\mathrm{bar}})}.
    \label{eqn:prominence}
\end{equation}

Central fields suffer from less contamination compared to those at the periphery. Therefore, we select stars from fields located within the region $|\ell|<0.9\degree$ and $-0.4\degree<b<0.25\degree$, which are highlighted in green in Fig.~\ref{fig:kmos_fields}. This selection is also motivated by the analysis of \PaperI, which used simulations to show that the vertex deviation of the nuclear component should dominate in this region (see their Figures 4 and 5). This selection reduces the sample to $1456$ stars. 

\begin{figure}
   \centering
    \includegraphics[width=.8\linewidth]{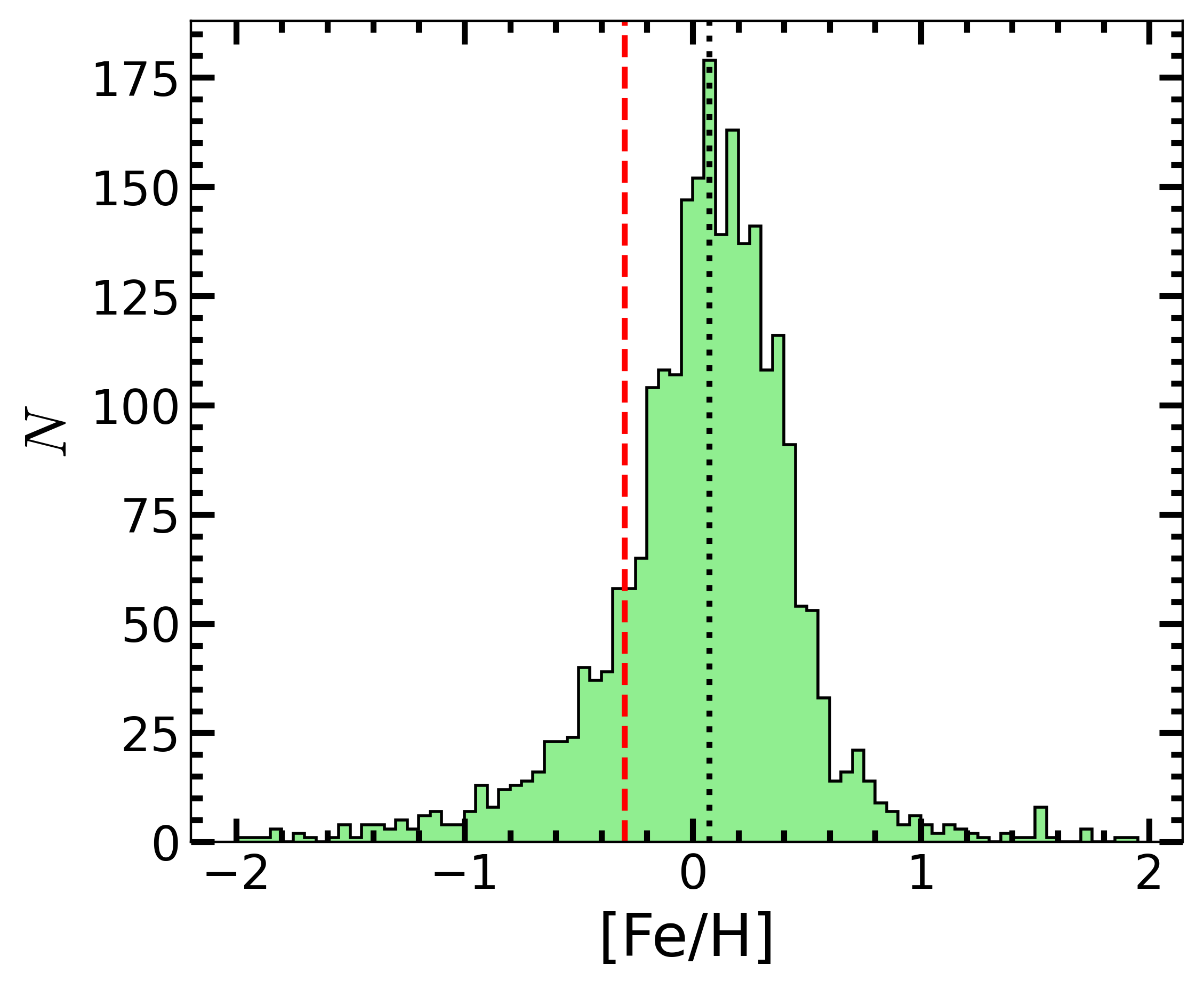}
    \caption{The \feh\ distribution of the entire KMOS dataset. The vertical dashed line at $\feh = -0.3$ marks the boundary between the metal-rich component, characterised by disc-like orbits, and the kinematically hotter, metal-poor component as shown in \citet{Nogueras+2024}. The median of the distribution is shown by the dotted black line.}
    \label{fig:kmos_feh}
\end{figure}

In Fig.~\ref{fig:kmos_feh}, we plot the \feh\ distribution of the entire sample. Based on the Gaussian Mixture Modelling of \citet{Nogueras+2024}, the NSD region appears to comprise two components: a metal-rich, disc-like population dominating the sample, and a metal-poor tail likely representing interlopers from the primary bar \citep[see also][]{Schultheis+2021}. To further minimise contamination from the primary bar population, we exclude stars with $\feh < -0.3$ (red dashed line), reducing our sample further to $1239$ stars

\begin{figure}
   \centering
    \includegraphics[width=\linewidth]{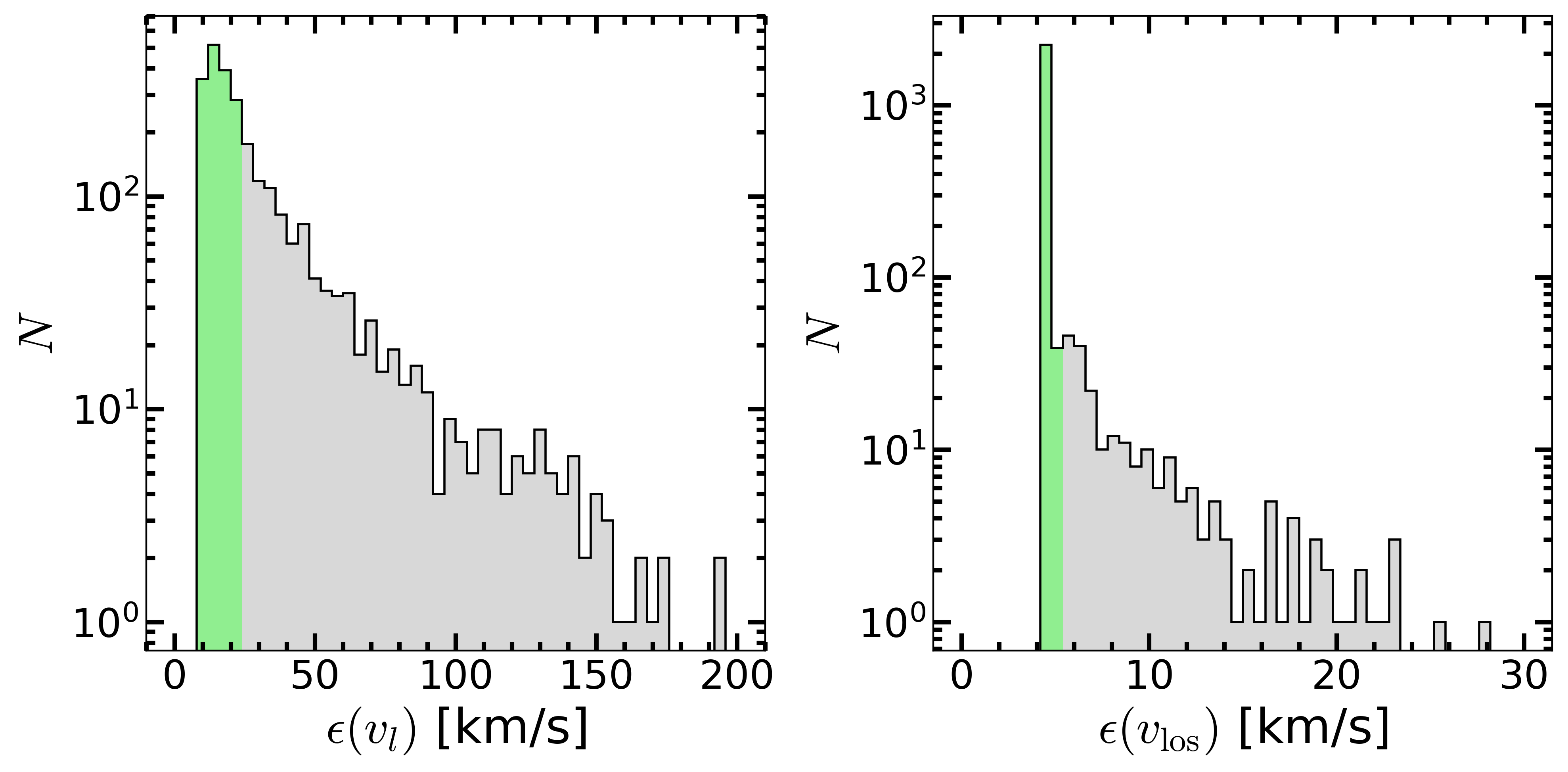}
    \caption{Histograms showing the error of the $v_{\ell}$ (left) and $v_{\mathrm{los}}$ (right) velocities. The gray sections indicate the stars which have been omitted from all samples in our analysis.}
    \label{fig:kmos_vels}
\end{figure}

We also make quality cuts on the velocities. In order to convert $\mu_\ell$ to $v_{\ell}$, we assumed a Galactocentric distance of $R=8.2\kpc$ \citep{Leung+2023} for all stars in our sample. The impact of this assumption on the results will be explored in Sec~\ref{sec:monte_carlo}. Fig.~\ref{fig:kmos_vels} shows the error distributions for $\mu_{\ell}$ (left) and $v_{\mathrm{los}}$ (right) for our KMOS+VIRAC2 dataset. We carry out quality cuts in absolute error of the velocities; $\epsilon(v_{\ell})<19\ {\mathrm{km/s}}$ (corresponding to $0.5\ {\mathrm {mas/yr}}$) and $\epsilon(v_{\mathrm{los}})<10\ {\mathrm{km/s}}$. In Fig.~\ref{fig:kmos_vels}, the cuts are reflected by the vertical dashed lines. Stars from the main bar typically reach larger heights above the mid-plane, and so are likely to have higher vertical proper motions \citep[e.g.,][]{Shahzamanian+2022}. Following \citet{Nogueras-Lara+2022}, therefore, we also require that $\mu_b < 5\, {\mathrm{mas/yr}}$. After applying these criteria, the dataset is reduced to 594 stars, which we denote as Sample~A.

We consider three additional samples for our analysis. Sample B is constructed by applying identical quality cuts on the velocities $\mu_{\ell}$ and $v_{\mathrm{los}}$, and includes stars located in the peripheral observational fields of the KMOS survey, indicated by gray markers in Fig.~\ref{fig:kmos_fields}. For this sample, we relax the metallicity constraint $\feh < -0.3$, thus incorporating stars across the full metallicity range. This results in a total of 505 stars, which we denote as Sample~B (or, equivalently, the bulge sample).

\begin{figure}
   \centering
    \includegraphics[width=.9\linewidth]{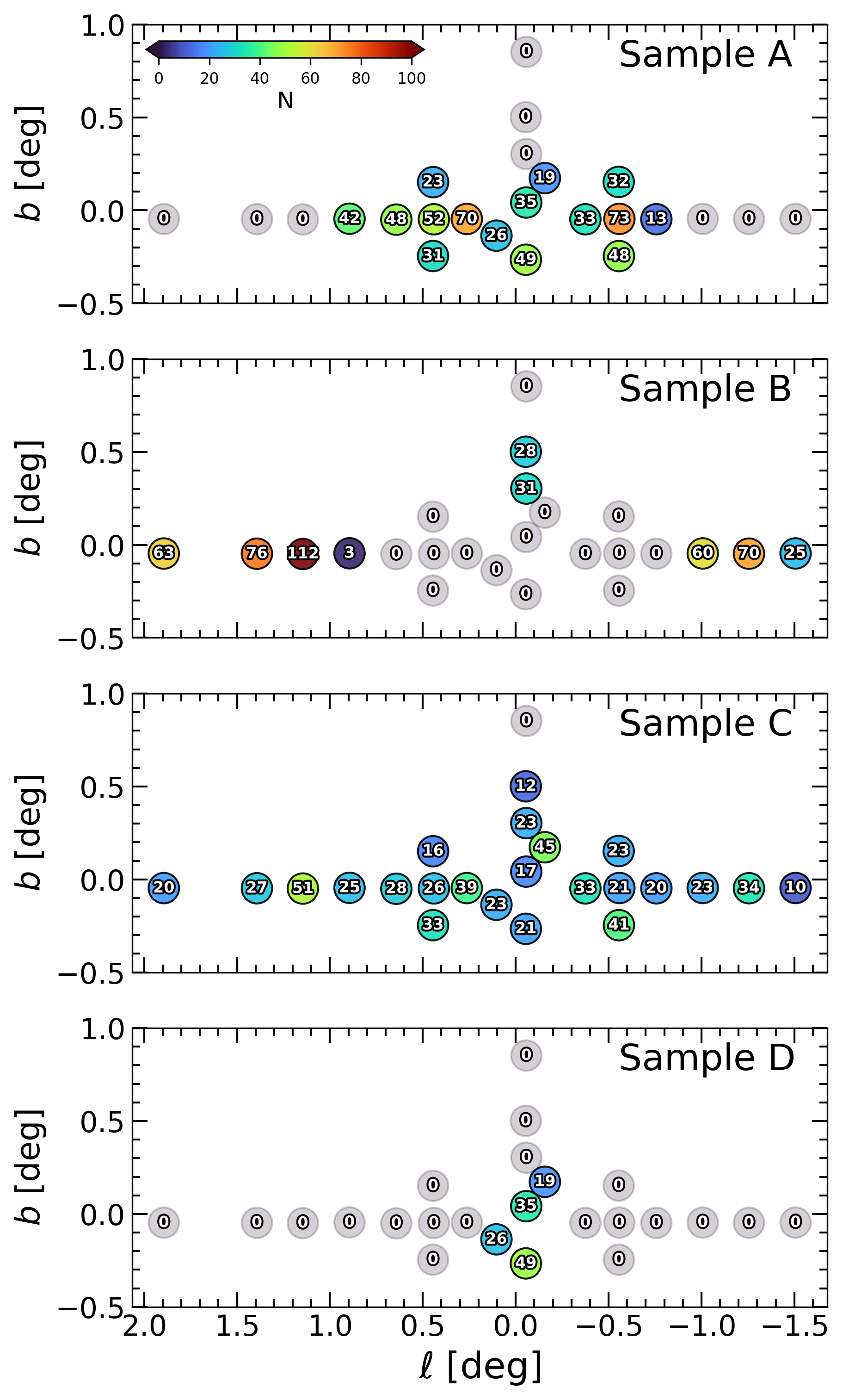}
    \caption{The fields from the KMOS survey of \citet{Fritz+2021} for each of the samples coloured by star count. We also annotate the star count in each individual field.}
    \label{fig:kmos_fields_no}
\end{figure}

We also apply the same quality cuts on velocities for Sample~C, which constitutes metal-poor ($[\mathrm{Fe/H}] < 0$) stars from all KMOS fields (except field 5), and comprises 611 stars. Lastly, sample D has all the same quality cuts of sample A, but only includes stars from the most central fields (1, 2, 7, 12 in Fig.~\ref{fig:kmos_fields}). In Fig.~\ref{fig:kmos_fields_no} we show the star count distribution in the KMOS fields for each of the samples (different rows) defined above.

\begin{figure*}
   \centering
    \includegraphics[width=\linewidth]{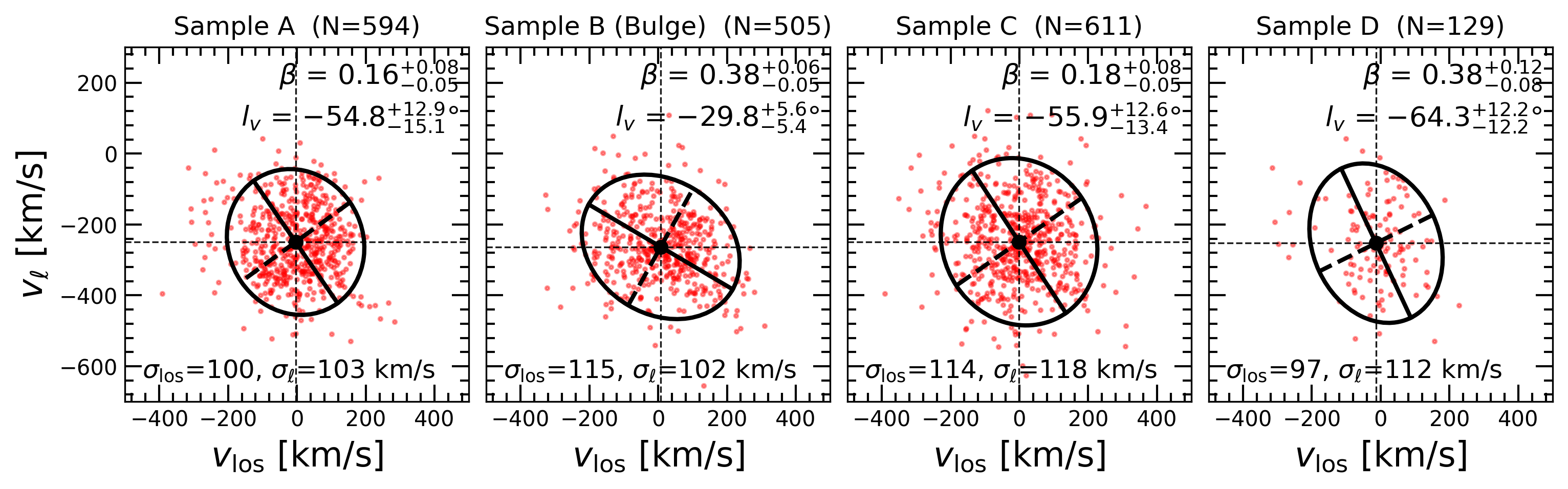}
    \caption{Velocity ellipse for all samples listed in Table~\ref{table:1}. The solid and dashed black lines indicate the major and minor axes of the velocity ellipse, respectively. The vertex deviation \lv\ and the anisotropy $\beta$ are annotated.}
    \label{fig:velocity_ellipse}
\end{figure*}

\section{Vertex deviation of the velocity ellipsoid}\label{sec:vertex_dev}

We compute the vertex deviation, \lv, and anisotropy parameter, $\beta$, of the $(v_{\ell}, v_{\mathrm{los}})$ velocity ellipsoid. The vertex deviation measures the angular offset between the major axis of the velocity ellipse and the nearest coordinate axis, and is defined as
\begin{equation}
    l_{\mathrm v}= \frac{1}{2}\arctan\left(\frac{2\sigma^{2}_{1 2}}{\sigma^{2}_{11}-\sigma^{2}_{22}}\right).
    \label{eqn:vertex_deviation}
\end{equation}
where $1$ and $2$ denote the $v_{\rm los}$ and $v_\ell$ directions respectively. By definition, \lv\ takes values $-90\degree \leq l_{\mathrm v} \leq 90\degree$. The anisotropy parameter is given by
\begin{equation}
    \beta= 1-\frac{\sigma^{2}_{\mathrm{min}}}{\sigma^{2}_{\mathrm{max}}},
    \label{eqn:beta}
\end{equation}
where $\sigma^{2}_{\mathrm{min}}$ and $\sigma^{2}_{\mathrm{max}}$ are the eigenvalues of the velocity-dispersion tensor, taking values $0 \leq \beta \leq 1$. Equation~\ref{eqn:vertex_deviation} measures the degree to which the local direction of highest dispersion deviates from alignment with the coordinate axes. In an axisymmetric system, the direction of maximum dispersion is expected to be aligned along the coordinate axes, with $\sigma^2_{\ell r}=0$, resulting in $l_{\mathrm v} = 0\degree$ or $l_{\mathrm v} = \pm 90\degree$. Conversely, the presence of bar-supporting orbits, which preferentially stream along the bar's major axis, leads to a non-zero covariance and hence a non-vanishing \lv. We refer the reader to \PaperI\ for a detailed derivation and discussion of these quantities.

Following the methodology of \citet{Fernandez+2025}, we estimate the uncertainties on $l_v$ and $\beta$ using a bootstrap resampling approach. For each sub-sample defined in this study, we resample the data with replacement over $B = 10^{4}$ iterations and recompute $l_v$ and $\beta$ for each resample, generating bootstrap distributions for both quantities. We quote the 16th and 84th percentile bounds of these distributions as the lower and upper uncertainty, respectively, giving asymmetric error bars on each measurement. We note that more isotropic velocity distributions typically yield larger uncertainties in $l_v$, as the orientation of the ellipse's major axis becomes increasingly poorly constrained.

\section{Results}\label{sec:results}

We apply the vertex deviation analysis on the subsamples defined in Sec.~\ref{sec:data}. Sample~A constitutes our primary sub-sample, comprised of stars from the central KMOS fields (green markers in Fig.~\ref{fig:kmos_fields}), where the contamination from main bar stars is expected to be $\lesssim 25\%$. The velocity ellipse for this sub-sample is displayed in the left panel of Fig.~\ref{fig:velocity_ellipse}, with the major and minor axes indicated by the solid and dashed lines, respectively. We measure a significant vertex deviation of $l_v = -54.8^{+13.1}_{-14.8}\,^\circ$ and an anisotropy parameter of $\beta = 0.16^{+0.08}_{-0.05}$. The velocity dispersions are $\sigma_{\mathrm{los}} = 100\,\mathrm{km\,s^{-1}}$ and $\sigma_\ell = 103\,\mathrm{km\,s^{-1}}$.

Sample~B (the "bulge" sample) comprises stars from the peripheral KMOS fields (gray markers in Fig.~\ref{fig:kmos_fields}), which are expected to suffer from high contamination by large-scale bar/bulge stars. The velocity ellipse is shown in the middle panel of Fig.~\ref{fig:velocity_ellipse}. As expected from a sample mainly composed of main bar/bulge stars, we measure a negative vertex deviation of $l_v = -29.8^{+5.5}_{-5.4}\,^\circ$ with high anisotropy $\beta = 0.38^{+0.06}_{-0.05}$. The velocity dispersions are $\sigma_{\mathrm{los}} = 115\,\mathrm{km\,s^{-1}}$ and $\sigma_\ell = 102\,\mathrm{km\,s^{-1}}$. In contrast to Sample~A, the line-of-sight direction has the highest dispersion, consistent with what is expected from the main bar oriented at $\approx 27^\circ$ \citep{Zhao+1996, Babusiaux+2010}.

Sample~C includes stars from all fields with $[\mathrm{Fe/H}] < 0$ (see third row of Fig.~\ref{fig:kmos_fields_no}). Since main bar stars are, on average, more metal-poor than NSD stars \citep{Schultheis+2021}, it suffers from very high contamination from main bar stars. Consequently, this results in a negative vertex deviation, $l_v = -55.9^{+12.6}_{-13.5}\,^\circ$. We also find a low anisotropy of $\beta = 0.18^{+0.08}_{-0.05}$.

Lastly, we examine the velocity ellipse for Sample~D, which is constituted only of stars in the innermost fields (1, 2, 7, 12). Similar to Sample~A, the vertex deviation is negative, $l_v = -64.3^{+12.1}_{-12.2}\,^\circ$, with notably higher anisotropy $\beta = 0.38^{+0.12}_{-0.07}$ than all other sub-samples. The direction of highest velocity dispersion is again in the Galactic longitude direction. The vertex deviations of Samples~A and~D thus differ from that of the bar-dominated Sample~B by $\approx 1.6\sigma$ and $\approx 2.6\sigma$, respectively. The above results are all summarised in Table~\ref{table:1}.

\begin{table*}
\caption{Summary of the samples defined in Sec.~\ref{sec:data} and their measured kinematic parameters. From left to right: sample name, sample size, metallicity range, fields included in the sample, vertex deviation $l_v$, anisotropy parameter $\beta$, line-of-sight and longitudinal velocity dispersions. Errors on $l_v$ and $\beta$ are the 16th and 84th percentile bounds from $10^4$ bootstrap resamples.}
\label{table:1}
\centering
\renewcommand{\arraystretch}{1.6}
\begin{tabular}{c c l l c c c c}
\hline\hline
Sample & $N$ & Metallicity & Fields & $l_v$ [deg] & $\beta$ & $\sigma_{\mathrm{los}}$ [km/s] & $\sigma_{\ell}$ [km/s] \\
\hline
A         & 594 & $\feh > -0.3$ & Central (green in Fig.~\ref{fig:kmos_fields})             & $-54.8^{+13.1}_{-14.8}$ & $0.16^{+0.08}_{-0.05}$ & 100 & 103 \\
B (Bulge) & 505 & All           & Peripheral (gray in Fig.~\ref{fig:kmos_fields} except 5)  & $-29.8^{+5.5}_{-5.4}$   & $0.38^{+0.06}_{-0.05}$ & 115 & 102 \\
C         & 611 & $\feh < 0$    & All except 5                                              & $-55.9^{+12.6}_{-13.5}$ & $0.18^{+0.08}_{-0.05}$ & 114 & 118 \\
D         & 129 & $\feh > -0.3$ & Very central (1, 2, 7, 12 in Fig.~\ref{fig:kmos_fields}) & $-64.3^{+12.1}_{-12.2}$ & $0.38^{+0.12}_{-0.07}$ &  97 & 112 \\
\hline
\end{tabular}
\end{table*}

\section{Distance and Velocity Uncertainties}\label{sec:monte_carlo}

\begin{figure}
   \centering
    \includegraphics[width=0.8\linewidth]{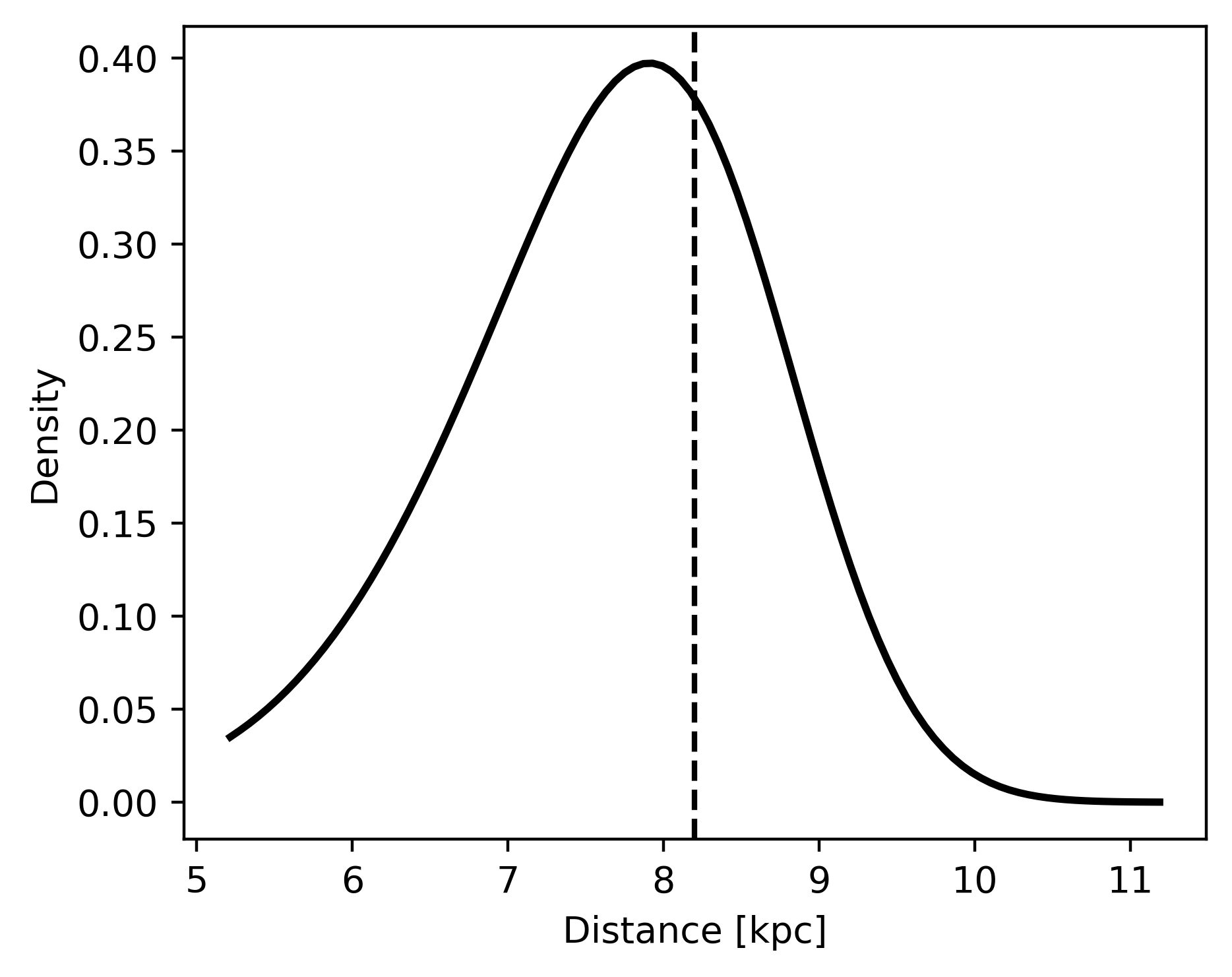}
    \caption{The skewed normal distribution from which we sample random Galactocentric distances for the Monte Carlo analysis. The distribution is centred at $\xi = 7.8$ kpc (close to our assumed distance of $R = 8.2$ kpc, marked by the dashed line) and is negatively skewed to reflect the higher likelihood of sampling stars on the near side of the disc.}
    \label{fig:distance_gauss}
\end{figure}

As outlined in Sec~\ref{sec:data}, we assumed that all stars lie at a Galactocentric distance of $R=8.2\kpc$ when converting $\mu_{\ell}$ to $v_{\ell}$. However, in reality stars will be distributed along the line of sight within the nuclear structure, with observational selection biasing our sample towards stars on the near side of the disc. To assess how this impacts our results, we model the distance distribution using a skew-Gaussian function\footnote{Using the scipy.stats.skewnorm() function} (shown in Fig.~\ref{fig:distance_gauss}) with a shape ("skewness") parameter of $-2$. While the stars in the samples likely have Galactocentric radii within a few hundred parsec of the assumed distance, we adopt a large standard deviation ($\sigma=1.5 \kpc$) of the skew-Gaussian function to account for the possibility that some stars may be located at larger Galactocentric radii. The distribution is skewed towards the near-side of the disc to reflect the higher likelihood of sampling nearby stars.

\begin{figure}
   \centering
    \includegraphics[width=\linewidth]{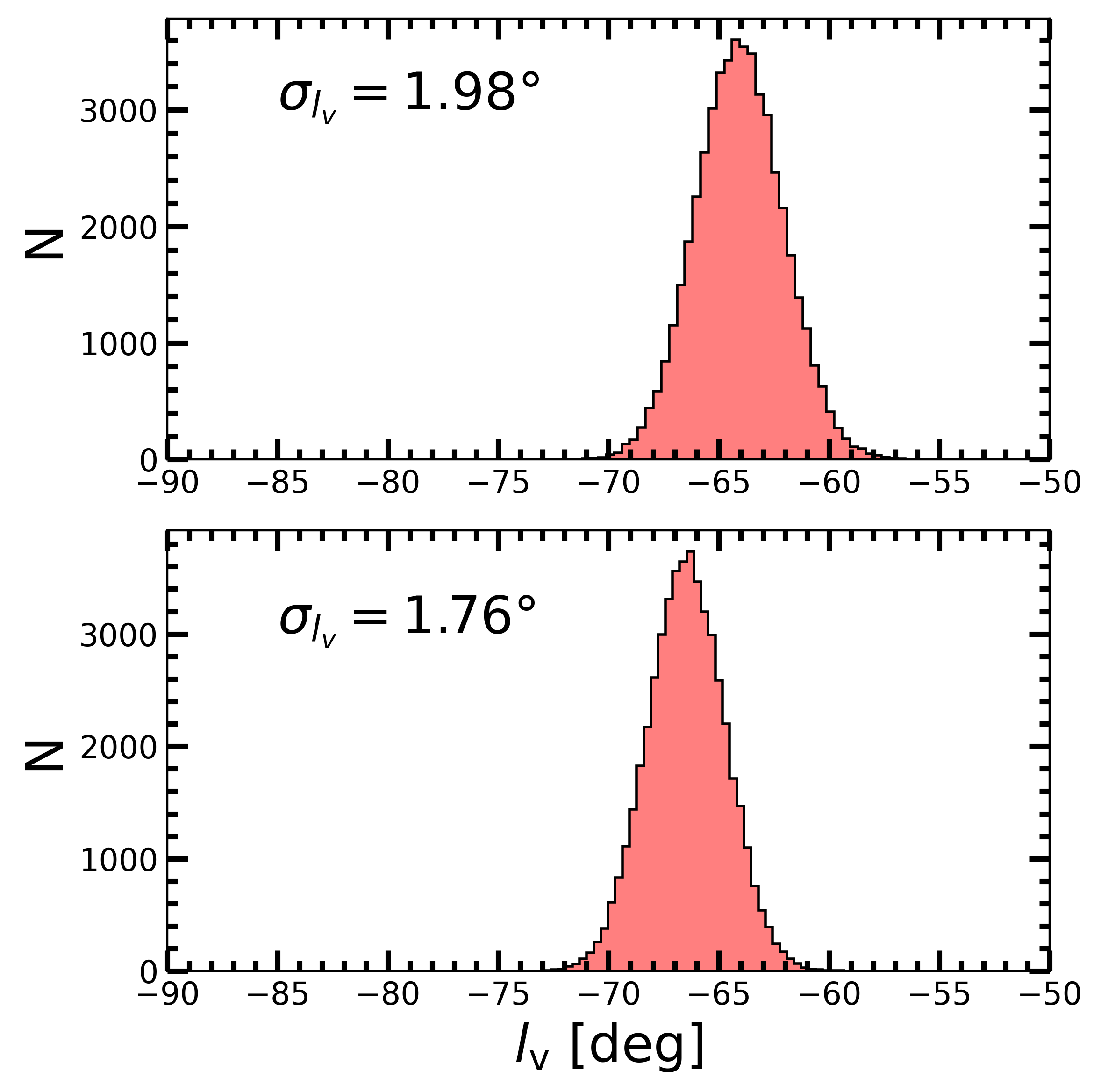}
    \caption{Results of 50,000-iteration Monte Carlo resampling with distance and velocity errors. The top panel shows the distribution of \lv\ when resampling distances from the skewed normal distribution, while the bottom panel shows the \lv\ distribution when resampling velocities within their measurement uncertainties. In both cases, the resulting uncertainties ($\sigma_{l_v}$) are negligible compared to the bootstrap resampling uncertainties.}
    \label{fig:lv_mc}
\end{figure}

For each individual star in sample A, we randomly sample a distance measurement from the skew-Gaussian in Fig.~\ref{fig:distance_gauss}, and re-calculate $v_{\ell}$. Once this is done for all stars in the sample, we re-compute \lv. We repeat the process for a total of $50,000$ iterations. The top panel of Fig.~\ref{fig:lv_mc} shows the distribution of \lv\ resulting from the Monte-Carlo resampling of distances. We find that the distribution has a standard deviation of $\sigma_{l_v} = 1.98\degree$, which is negligible when compared to the uncertainties associated with the observational samples when bootstrapping is applied (Sec.~\ref{sec:vertex_dev}).

We also assess how the uncertainties in $v_{\mathrm{los}}$ and $v_{\ell}$ impact measurement of \lv\ by propagating the errors through a Monte-Carlo resampling. For each star in the cleaned sample A, we generate two independent Gaussian distributions with means equal to the observed $v_{\mathrm{los}}$ and  $v_{\ell}$ and standard deviations equal to their respective measurement errors. The velocity component for each star is resampled for 50,000 iterations from its Gaussian, yielding a perturbed velocity set. For every iteration, we recompute \lv, thereby obtaining its distribution.

The bottom panel of Fig.~\ref{fig:lv_mc} shows the distribution of \lv\ resulting from the Monte-Carlo velocity resampling. The distribution is strongly peaked at $l_{\mathrm v} \approx -64\degree$ with a negative tail extending up to $l_{\mathrm v} \approx -68\degree$. The standard deviation is $\sigma_{l_v} = 1.76\degree$, again indicating that the errors arising from velocity uncertainties are negligible compared to the sampling uncertainties inferred from the bootstrap method.

\section{Impact of extinction on vertex deviation}\label{sec:extinction}

The Galactic centre suffers from severe and spatially non-uniform extinction, with significant variations occurring even within individual KMOS pointings (see Fig.~\ref{fig:ext_map}). To assess how extinction impacts our results, we adopt a two-stage de-reddening approach, prioritising individual stellar extinction estimates where possible and falling back to map-based values where necessary.

\subsection{Extinction estimation}\label{sec:ext_estimation}

\begin{figure}
   \centering
    \includegraphics[width=\linewidth]{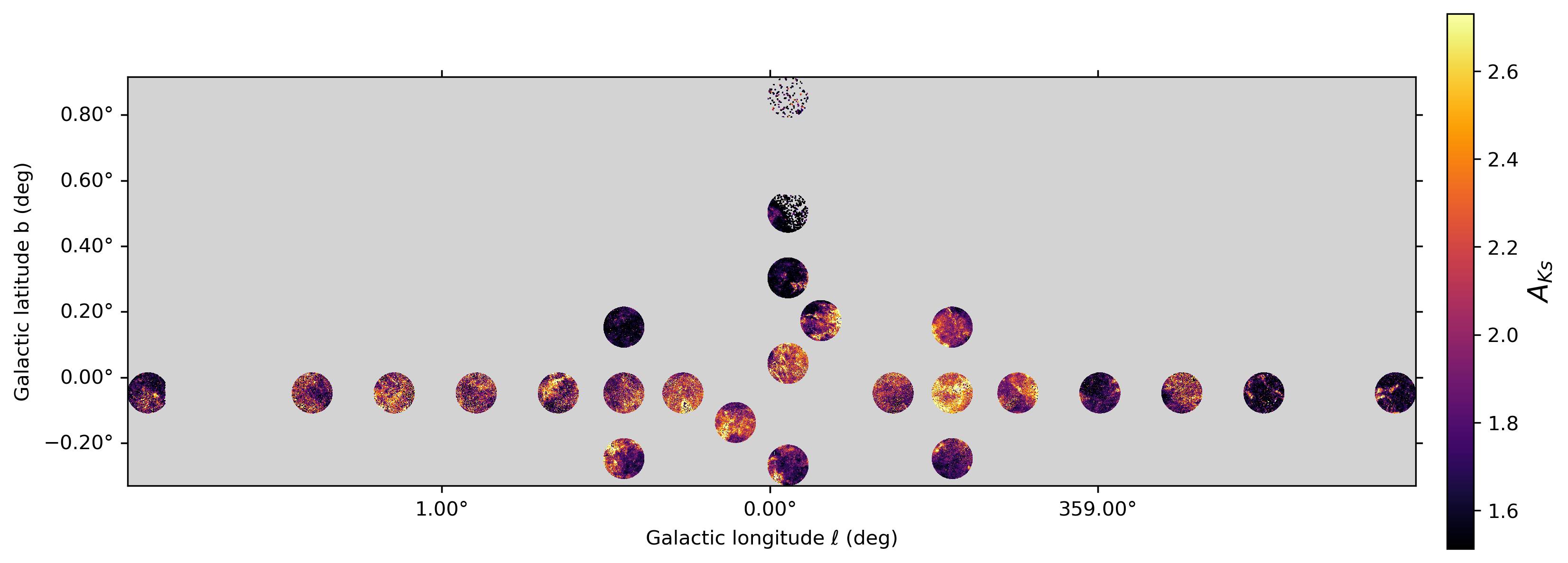}
    \caption{Spatial variation of $A_{K_s}$ extinction within each KMOS field, derived from the VIRAC2-based maps. Each field is coloured by the local extinction value, revealing the significant small-scale structure present even within individual pointings, as well as the broader gradient toward the Galactic centre.}
    \label{fig:ext_map}
\end{figure}

\begin{figure}
   \centering
    \includegraphics[width=\linewidth]{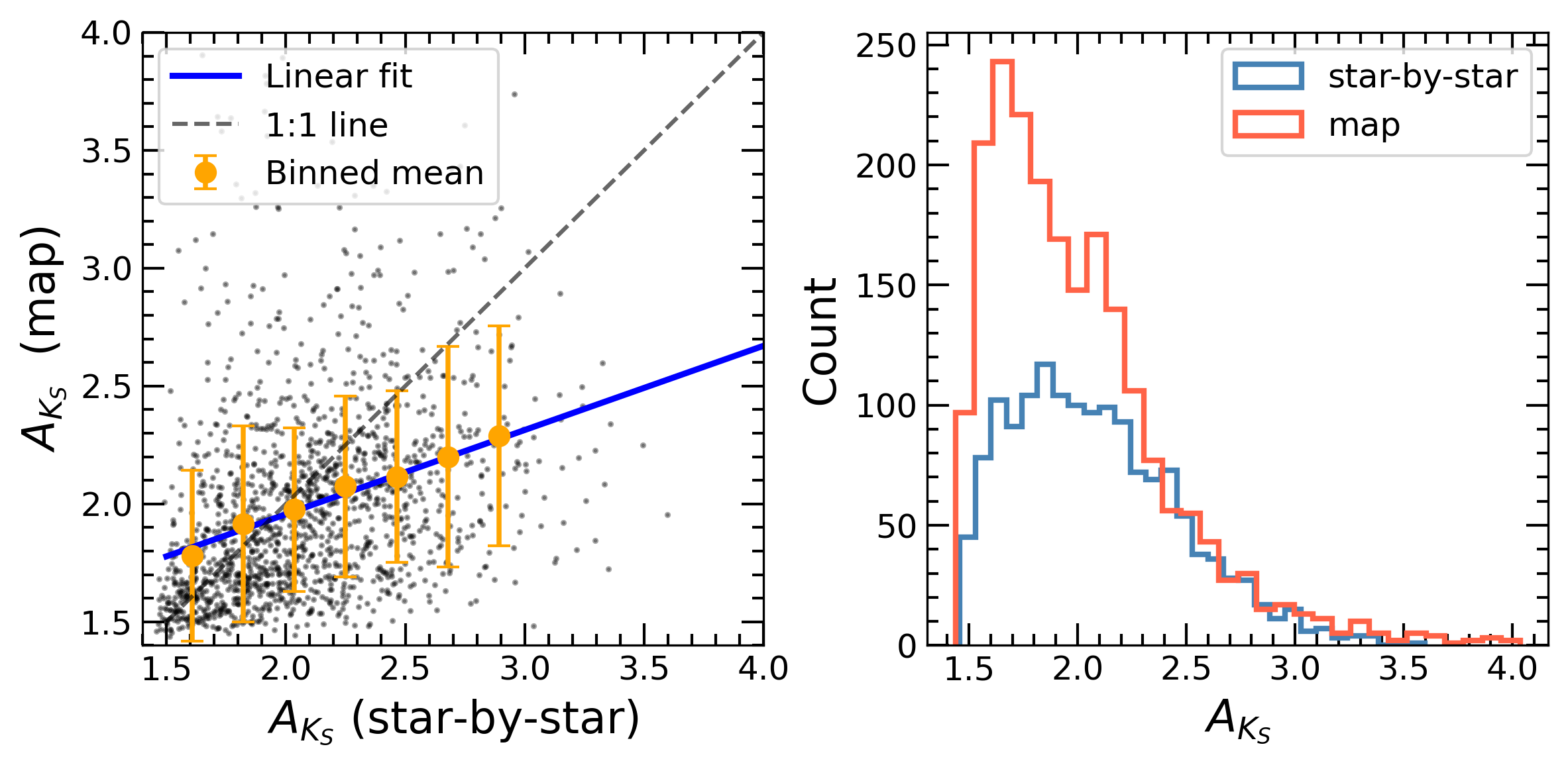}
    \caption{Comparison of extinction estimates for the full sample derived from the star-by-star and map-based methods. \emph{Left:} One-to-one comparison of the two extinction estimates for individual stars. The gray and blue lines show the 1:1 relation and a linear fit to the data, respectively. Orange markers indicate the binned mean extinction, with error bars reflecting the uncertainty in each bin. \emph{Right:} Distribution of $A_{K_s}$ values for the full sample derived from the star-by-star (blue) and map-based (orange) methods. The two approaches show good agreement for $A_{K_s} \lesssim 2.5$, with larger discrepancies confined to the high-extinction tail where only a small number of stars reside.}
    \label{fig:extinction_comp}
\end{figure}

\noindent\textit{Star-by-star extinction.} Our primary method computes individual extinction values for each KMOS target following the approach described in \citet{Nogueras-lara+2021}. A reference population of red giant stars is selected from the VIRAC2 CMD using $K_s \in [13.75, 16.75]$ and $H\mhyphen K_s \in [1.3, 3.5]$. These stars have sufficiently uniform intrinsic $H\mhyphen K_s$ colours ($(H\mhyphen K_s)_0 = 0.10 \pm 0.01$; \citealt{Nogueras-lara+2021}) to serve as reliable extinction tracers: any observed colour excess above this intrinsic value is attributed to dust. The extinction in the $K_s$ band for each reference star is computed via
\begin{equation}
    A_{K_s} = \frac{(H-K_s)_{\mathrm{obs}} - (H-K_s)_0}{(A_H/A_{K_s}) - 1},
    \label{eqn:nice_ext}
\end{equation}

\noindent where $(H-K_s)_{\mathrm{obs}}$ is the observed colour of the reference star, $(H-K_s)_0 = 0.10$ is the intrinsic red giant colour, and $A_H/A_{K_s} = 1.84 \pm 0.03$ is the extinction law derived for the Galactic centre by \citet{Nogueras-lara+2020b}. The extinction of each KMOS target is then estimated as an inverse-distance-weighted (IDW) mean of the surrounding reference star extinctions,
\begin{equation}
    A^{\mathrm{target}}_{K_s} = \frac{\sum^n_{i=1} d_i^{-p}\, A_{K_s,i}}
    {\sum^n_{i=1} d_i^{-p}},
    \label{eqn:idw}
\end{equation}

\noindent where $d_i$ is the angular distance from the target to reference star $i$, $A_{K_s,i}$ is its extinction from Eq.~\ref{eqn:nice_ext}, and $p = 0.25$ is the distance weighting exponent \citep{Nogueras-lara+2021}. An additional colour-consistency constraint is applied to the reference sample where only stars with $H\mhyphen K_s$ colours sufficiently close to that of the target are included, which implicitly selects stars at similar line-of-sight depths and avoids averaging over reference stars located in distinct extinction layers along the line of sight. A minimum of five reference stars satisfying both the proximity and colour-consistency criteria is required; targets for which this threshold is not met are assigned extinction values from the map-based method described below. We regard this as the most reliable method for individual stellar de-reddening.

\noindent\textit{Extinction maps.} For stars that could not be assigned an individual extinction estimate we instead use extinction maps constructed for each KMOS pointing following \citet{Nogueras-lara+2021}. Using the same red giant reference population, Eq.~\ref{eqn:nice_ext} is first applied to each reference star to obtain its individual $A_{K_s}$. A map is then constructed on a regular grid at $2\arcsec$ resolution: for each pixel, Eq.~\ref{eqn:idw} is applied using the surrounding reference stars within a fixed search radius, without the colour-consistency constraint used in the star-by-star method. A minimum of five reference stars within the search radius is required for a pixel to be assigned a value. The extinction at any target position is then read off from this map. The significant small-scale structure present even within individual pointings, as well as the broader gradient toward the Galactic centre, are illustrated in Fig.~\ref{fig:ext_map}. As shown in Fig.~\ref{fig:extinction_comp}, the two approaches agree well for $A_{K_s} \lesssim 2.5$, with mild systematic differences at higher extinctions. In our samples, $32\%$ of the 594 stars in sample A and $18\%$ of the 129 stars in sample D required map-based extinction estimates, with mean $A_{K_s} \leq 2.1$ in both cases --- well within the regime of good agreement between the two methods. The choice of fallback therefore has a negligible impact on the bulk of the de-reddened sample.

\subsection{Effect on the vertex deviation and anisotropy}\label{sec:ext_vertex}

\begin{figure*}
\centering
\includegraphics[width=.8\linewidth]{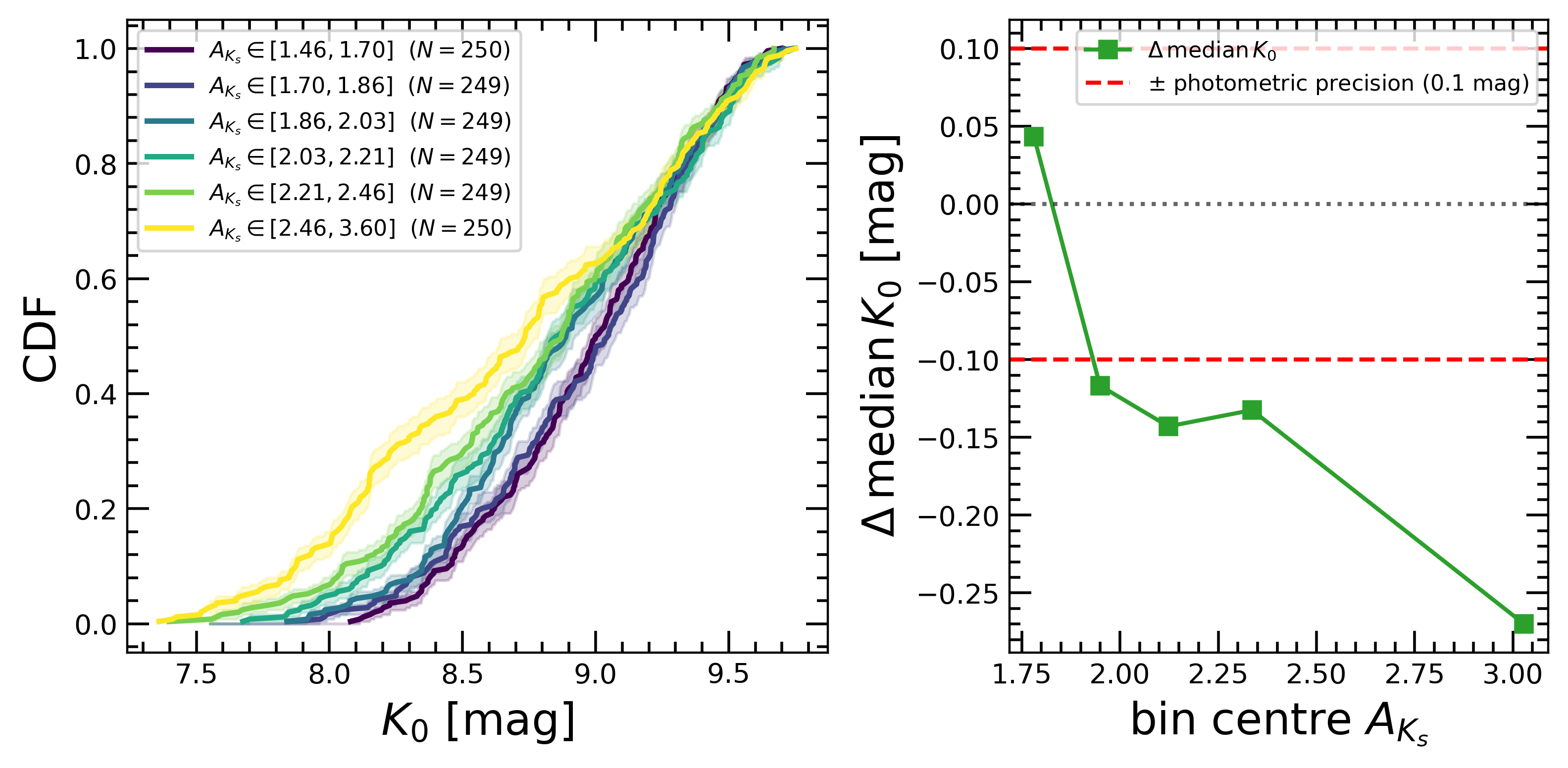}
\caption{Assessment of extinction-driven incompleteness in the kinematic sample. \emph{Left:} Cumulative distribution functions of the de-reddened magnitude $K_0$ in equal-population quantile bins of $A_{K_s}$ (coloured curves, see legend), with shaded $16$--$84$ percentile bootstrap envelopes. The lowest-$A_{K_s}$ bin serves as the unbiased reference. \emph{Right:} Difference between the median $K_0$ of each bin and that of the reference bin, plotted at the bin centre. Dashed lines mark $\pm 0.1$~mag, the typical photometric precision of the input catalogue; shifts within this band are consistent with the per-star measurement noise floor. Negative values indicate bins whose stars are typically brighter than the reference, the expected signature of either a faint-end magnitude limit removing faint stars at high $A_{K_s}$ or a bright-end saturation limit removing bright stars at low $A_{K_s}$.}
\label{fig:completeness}
\end{figure*}

While extinction does not directly affect the line-of-sight velocities or proper motions used to construct the velocity ellipse, it enters our analysis through the photometric selection that defines the kinematic sample. The \citet{Fritz+2021} selection was applied in the observed colour--magnitude plane (Fig.~\ref{fig:selection}), and is therefore sensitive to the spatially-varying extinction across the KMOS pointings. To assess whether this biases our measurements, we perform two complementary tests on the sample itself.

We emphasise that these tests are specifically designed to identify extinction-dependent biases, that is, biases whose magnitude varies with $A_{K_s}$ across the field of view. The absolute completeness of the underlying KMOS+VIRAC2 spectroscopic selection within the \citet{Fritz+2021} CMD box is much lower than unity, with only $\sim$few percent of photometrically eligible stars in each field actually targeted. This overall undersampling is set by the survey's multiplexing rather than by any astrophysical property of the targets, and we assume it is approximately uniform within each field.

We first assess whether the sample is photometrically complete in the de-reddened $K$-band magnitude, $K_0 = K_s - A_{K_s}$, across the range of extinctions probed. We divide the sample into six equal-population quantile bins in $A_{K_s}$ and compare the $K_0$ distribution in each bin to that of the lowest-extinction reference bin (Fig.~\ref{fig:completeness}). The cumulative distribution functions drift visibly toward brighter $K_0$ at higher $A_{K_s}$ (left panel), with the highest-$A_{K_s}$ bin showing minimal overlap with the bootstrap envelope of the reference. The shifts in the median $K_0$ relative to the reference (right panel) confirm this quantitatively: for $A_{K_s} \gtrsim 1.9$, the median $K_0$ is systematically brighter than the reference by $0.13$--$0.27$~mag, exceeding the typical photometric precision of the input catalogue ($\sim 0.1$~mag). This signature originates in the VIRAC2 saturation limit rather than in the KMOS faint-end magnitude limit. \citet{Fritz+2021} report that the KMOS observational depth in the NSD region only begins to truncate the underlying luminosity function at $A_{K_s} \approx 5$; since our sample reaches only $A_{K_s} \approx 3.5$, the KMOS magnitude limit is not the source of the observed shift. The VIRAC2 photometric catalogue, however, saturates at $K_s \approx 10$~mag, which means that in fields with comparatively low extinction the intrinsically brightest NSD stars ($K_0 \approx 6.5$--$8.5$) reach $K_s < 10$ and are excluded by the saturation cut, while the same stars in fields of higher extinction are reddened to $K_s > 10$ and survive. The net effect is that the low-$A_{K_s}$ bins are missing the bright end of the underlying NSD luminosity function, while the high-$A_{K_s}$ bins retain it — producing the monotonic shift of the median $K_0$ towards brighter values with increasing $A_{K_s}$ seen in Fig.~11. The bias is moderate over $A_{K_s} \in [2.0, 2.5]$ and becomes severe at $A_{K_s} > 2.5$.

\begin{figure}
\centering
\includegraphics[width=.9\linewidth]{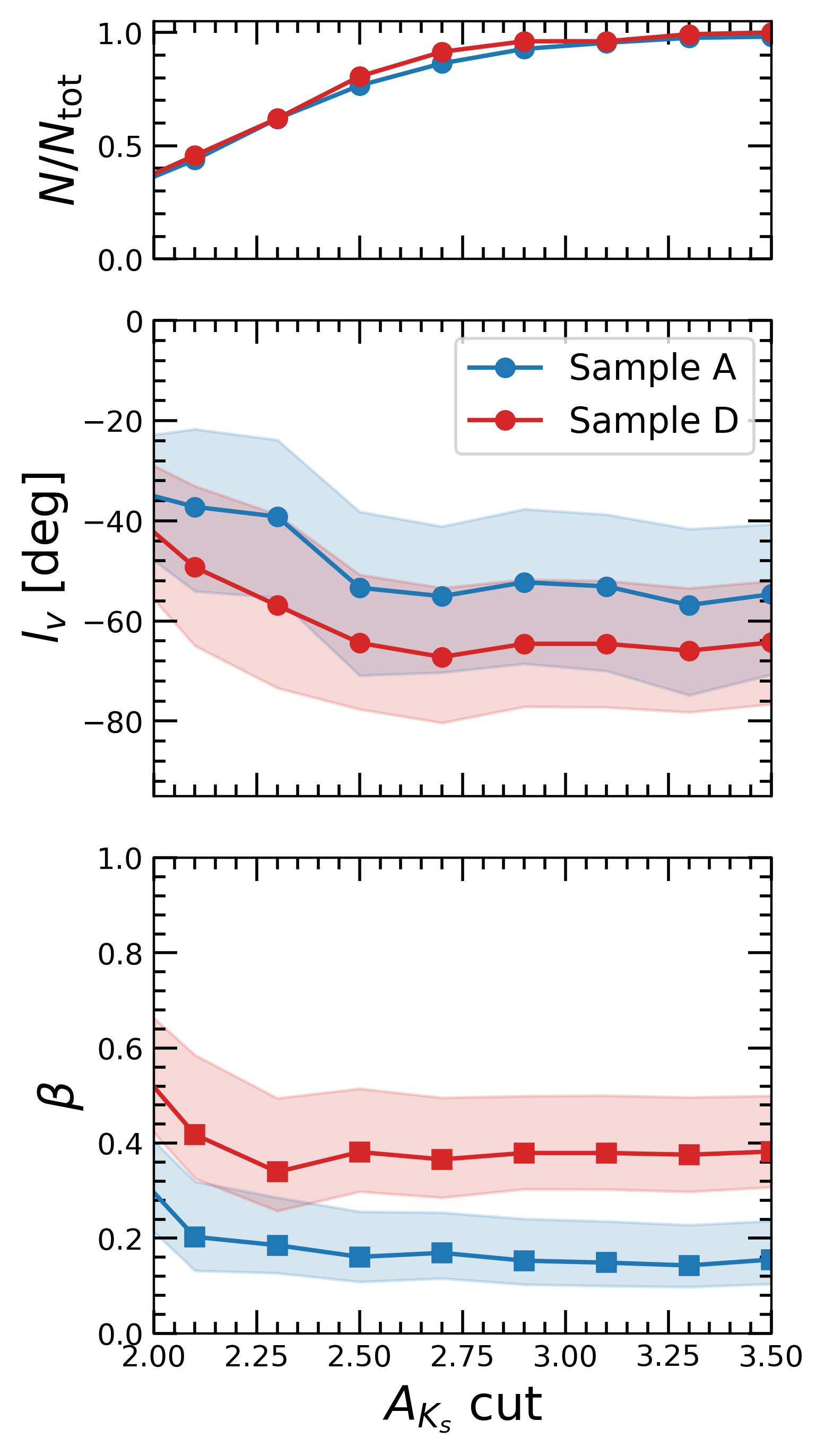}
\caption{Robustness of the kinematic measurements to extinction-driven incompleteness. \emph{Top:} fraction of stars surviving an upper cut on $A_{K_s}$ for sample A (blue), and sample D (red). \emph{Middle:} vertex deviation $l_v$ as a function of the same cut, with shaded bands showing the bootstrap uncertainty. \emph{Bottom:} same as the middle panel but for the anisotropy parameter, $\beta$. At the loosest cut, samples A and D reproduce the values reported in Table~\ref{table:1}; both quantities remain stable within their bootstrap uncertainties as the cut is tightened down to $A_{K_s} < 2.0$, below which the sample size becomes insufficient for a reliable measurement.}
\label{fig:vdev_ext}
\end{figure}

We then test whether this photometric incompleteness propagates into our kinematic measurements. For samples A and D we recompute $l_v$ and $\beta$ on subsamples defined by progressively stricter upper cuts in $A_{K_s}$ (Fig.~\ref{fig:vdev_ext}). At the loosest cut, both samples reproduce the values reported in Table~\ref{table:1}. As the cut is tightened from $A_{K_s} < 3.5$ down to $A_{K_s} < 2.0$, both $l_v$ and $\beta$ remain consistent within their bootstrap uncertainties for both samples, despite the median $K_0$ shift varying from $-0.27$~mag to $-0.13$~mag over the same range. Below $A_{K_s} \approx 2.0$ the bootstrap uncertainties begin to grow as the sample sizes shrink ($N/N_\mathrm{tot} \lesssim 0.4$ for Samples~A and~D), and we restrict our quantitative robustness statement to the regime above this value. We additionally note that the surviving sample at strict upper $A_{K_s}$ cuts is increasingly weighted towards low-extinction fields, where the foreground-rejection cut introduced in Sec.~\ref{sec:data} is least clean: in fields with comparatively low extinction, NSD stars are themselves less reddened and may have observed $(H-K_s)$ colours close to or below the $1.3$ mag threshold, leading to their inadvertent removal alongside true foreground interlopers. This combination of small samples and a target-selection bias whose impact grows precisely in this regime precludes a reliable kinematic test below $A_{K_s} \approx 2.0$ with the present data.

We additionally note that the Galactic-centre extinction is well known to be asymmetric in Galactic longitude, with approximately two-thirds of the dust within $|\ell| < 1.5^\circ$ concentrated at positive longitudes \citep{Bally+1988, Henshaw+2023}. In \PaperI\ (their Sec.~3.4 and Fig.~11) we constructed a toy selection function reproducing this longitude-dependent extinction and applied it to an axisymmetric NSD model, finding that $l_v$ is then shifted towards positive values, in the opposite direction to our negative measurement. The longitude asymmetry of the Galactic-centre extinction is therefore unable to account for the observed signal.

We conclude that, although a saturation-driven photometric incompleteness is detectable in the sample for $A_{K_s} \gtrsim 1.9$, the resulting selection bias does not propagate into our kinematic diagnostics. The vertex deviation and anisotropy measurements reported in Sec.~\ref{sec:results} are therefore decoupled from the extinction-driven incompleteness, and the dominant uncertainty in $l_v$ remains the statistical sampling uncertainty characterised by the bootstrap approach of Sec.~\ref{sec:vertex_dev}.

\section{Impact of bar contamination on vertex deviation}\label{sec:contamination}

The kinematic samples used in this work are drawn from within $|\ell| < 0.9^\circ$, $-0.4^\circ < b < 0.25^\circ$, where the NSD is expected to dominate the surface density. They are nonetheless expected to retain some level of contamination from primary-bar/bulge interlopers along the line of sight. To quantify the impact of this contamination on our vertex deviation measurement, we combine the predictions of MW-tailored simulations with two complementary sample-based tests on the data themselves.

\subsection{Constraints from MW-calibrated simulations}\label{sec:contamination_paperI}

The level of bar/bulge contamination in the central KMOS fields is constrained by independent observational and dynamical analyses to be of order $20$--$25\%$ \citep{Sormani+2022, Nogueras+2024}. In \PaperI\ we constructed a simulation framework specifically tailored to the MW: the NSD is the self-consistent dynamical model of \citet{Sormani+2022}, fitted directly to the same KMOS+VIRAC2 kinematic data used in the present work. The large-scale bar is the N-body model of \citet{Deg+2025}, rescaled to match the observed bar pattern speed and semi-major axis \citep{Wegg+2015, Portail+2017}; and the NSD/bar surface-density ratio is fixed to that of \citet{Sormani+2022} so that the contamination level at every $(\ell, b)$ position in the model matches what is observed. In this MW-calibrated framework, an axisymmetric NSD seen through the expected $20$--$25\%$ bar contamination still produces $l_v$ close to $\pm 90^\circ$ within the central region $|\ell| < 0.9^\circ$, $|b| < 0.25^\circ$ (their Fig.~5 and Sec.~3.2). At the contamination level inferred for the MW, primary-bar interlopers do not by themselves displace $l_v$ to the strongly negative values measured in samples~A and~D.

\subsection{Sample-based tests}\label{sec:contamination_data}

The four samples defined in Sec.~\ref{sec:data} probe the kinematic signal at different expected levels of bar contamination, providing a self-consistent sample-based test of the contamination dependence. Sample~B, dominated by primary-bar and bulge stars from the peripheral fields, yields $l_v = -29.8^{+5.5}_{-5.4}\,^\circ$. Samples~A and~D, which use the central fields and the metallicity cut $[\mathrm{Fe/H}] > -0.3$ designed to suppress the metal-poor bar contribution \citep{Schultheis+2021, Nogueras+2024}, yield $l_v = -54.8^{+13.1}_{-14.8}\,^\circ$ and $l_v = -64.3^{+12.1}_{-12.2}\,^\circ$, respectively. The progression is monotonic in the direction expected if reducing the bar contamination drives $l_v$ towards the underlying nuclear-component value, and the difference between Samples~A/D and~B is statistically significant ($\approx 1.6\sigma$ and $\approx 2.6\sigma$, respectively).

\begin{figure}
\centering
\includegraphics[width=.9\linewidth]{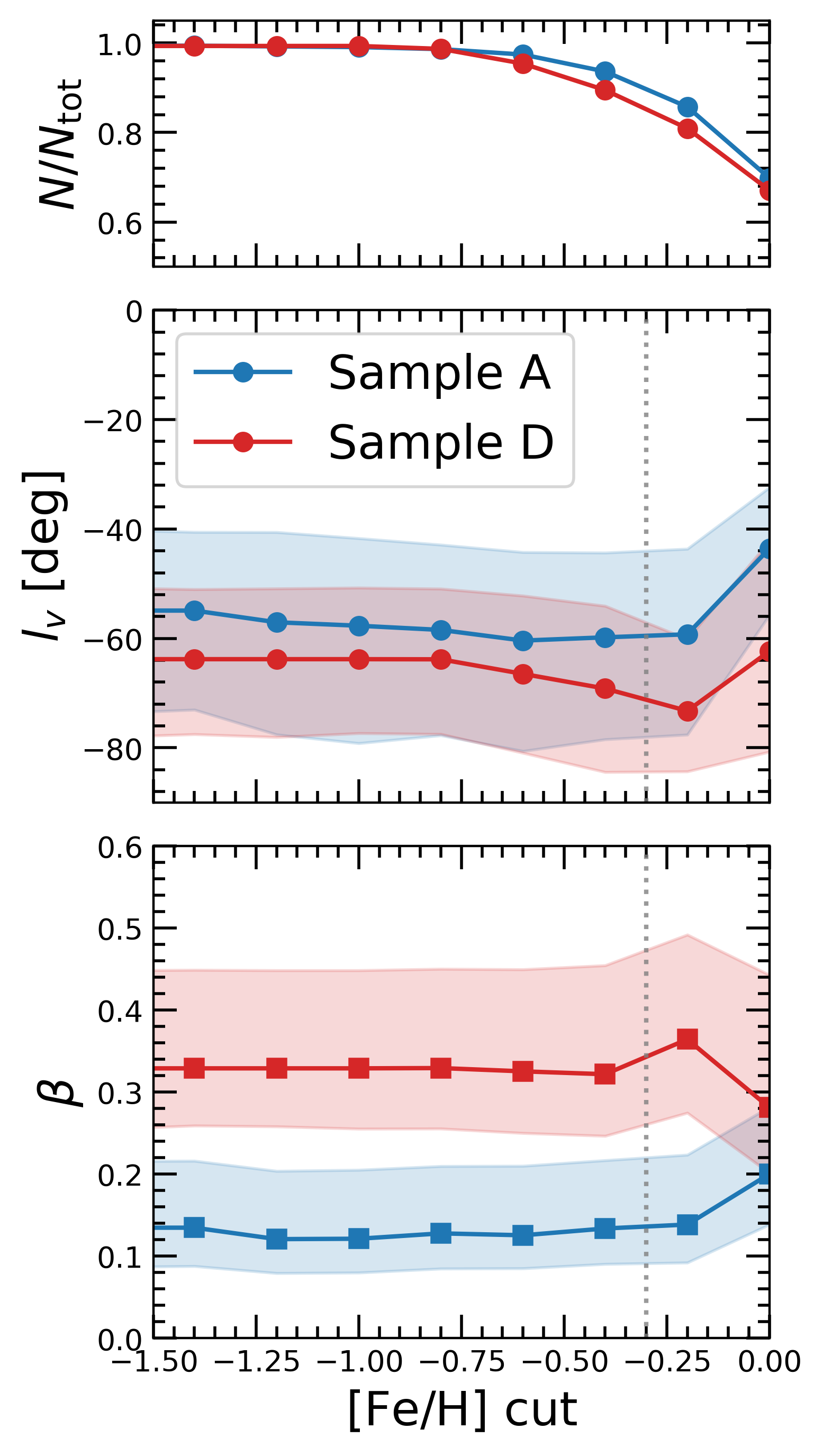}
\caption{Robustness of the kinematic measurements to the metallicity cut. \emph{Top:} fraction of stars surviving a lower cut on $[\mathrm{Fe/H}]$ for sample~A (blue) and sample~D (red). \emph{Middle:} vertex deviation $l_v$ as a function of the same cut, with shaded bands showing the $1\sigma$ bootstrap uncertainty. \emph{Bottom:} same as the middle panel but for the anisotropy parameter $\beta$. The dotted vertical line indicates the baseline cut $[\mathrm{Fe/H}] > -0.3$ used to define samples~A and~D. Both quantities are stable within their bootstrap uncertainties across the full swept range, including at the loosest cut where contamination from metal-poor bar interlopers is highest.}
\label{fig:vdev_Fe}
\end{figure}

We further tested whether the $l_v$ signal in samples~A and~D is sensitive to the specific metallicity threshold used to suppress bar contamination. Following the Gaussian-mixture decomposition of \citet{Schultheis+2021} and \citet{Nogueras+2024}, in which the metal-rich peak ($[\mathrm{Fe/H}] \gtrsim 0$) traces the NSD and the metal-poor tail traces the bar/bulge interlopers, we swept the lower threshold across $[\mathrm{Fe/H}]_{\min} \in [-1.5, 0.0]$, the range over which the cut progressively removes bar contaminants without yet eroding the metal-rich NSD peak, and recomputed $l_v$ and $\beta$ at each cut in steps of $0.05$ (Fig.~\ref{fig:vdev_Fe}). Both quantities are consistent with the values reported in Table~\ref{table:1} within their bootstrap uncertainties across the full swept range. Crucially, loosening the cut to $[\mathrm{Fe/H}]_{\min} = -1.5$, which admits a larger fraction of metal-poor stars and therefore a higher level of expected bar contamination, leaves $l_v$ unchanged ($l_v = -54.8^\circ$ for sample~A and $-64.3^\circ$ for sample~D). A small drift in both samples is visible only at $[\mathrm{Fe/H}]_{\min} \to 0.0$, where the cut becomes aggressive enough to remove genuine NSD members from the metal-rich peak of the distribution; even there, the result remains within the bootstrap envelope. We additionally note that the absence of any drift towards the bar value of $l_v \approx -29.8^\circ$ at the loosest cuts is consistent with the metal-poor population in the central fields itself containing a non-trivial fraction of NSD members rather than being dominated by bar interlopers, in line with the GMM decomposition of \citet{Nogueras+2024}.

Taken together, the MW-calibrated simulations of \PaperI, the four-sample empirical sequence, and the metallicity-cut robustness test provide complementary, mutually consistent evidence that the strongly negative $l_v$ measured in samples~A and~D arises from a kinematically-distinct nuclear component rather than from primary-bar contamination. We note nonetheless that the available sample sizes do not yet allow us to fully exclude residual contributions from contamination at the few-percent level, and that this regime can only be tested directly with the larger samples expected from upcoming surveys (Sec.~\ref{sec:future_prospects}).

\section{Discussion and Conclusion}\label{sec:summary}

In this paper, we have characterised the ($v_{\mathrm{los}}, v_\ell$) velocity ellipse in the Galactic NSD region using subsamples from KMOS+VIRAC2. For the subsamples comprised of stars from central fields (samples A and D), we find a vertex deviation that is inconsistent both with that of the large-scale bar and with an axisymmetric NSD. For Sample~A, we measure $l_v = -54.8^{+13.1}_{-14.8}\,^\circ$ with an anisotropy of $\beta = 0.16^{+0.08}_{-0.05}$, which shows a separation of $\approx 1.6\sigma$ from Sample~B. For Sample~D, we measure a vertex deviation of $l_v = -64.3^{+12.1}_{-12.2}\,^\circ$ with an anisotropy of $\beta = 0.38^{+0.12}_{-0.07}$, representing a $\approx 2.6\sigma$ difference from Sample~B. Taken together, these measurementsprovide the first kinematic hints that the MW hosts a nuclear bar.

\subsection{Elliptical nuclear disc or nuclear bar?}

External galaxies often host NSDs that are not perfectly axisymmetric, but rather slightly elliptical \citep[e.g.,][]{Bittner+2020}, with the NSD's major axis oriented orthogonally to the primary bar \citep[e.g.,][]{Cole+2014}. Could the kinematic signature we observe arise from an elliptical NSD oriented orthogonally?

The MW's primary bar is oriented at about $27\degree$ to the Solar-GC line, with its near side at positive longitudes \citep{Wegg+2015}. An elliptical orthogonally oriented NSD would therefore have its near side oriented toward negative longitude. In \PaperI, we demonstrated that such a configuration produces a mildly positive vertex deviation. In contrast, our observed \lv\ is negative. Our measured \lv\ is therefore inconsistent with being due to an elliptical NSD.

\subsection{Implications for the Milky Way}

The presence of a nuclear bar would fundamentally alter the gravitational potential of the inner $R < 300\pc$ and hence, the orbital structure of stars and gas in this region \citep{Schultheis+2025}. This would require a significant re-evaluation of current theoretical models for the three-dimensional structure of the central molecular zone (CMZ) \citep{Henshaw+2023}, necessitating updates to our understanding of the gas kinematics and star formation in the Galactic centre. Moreover, the changed orbital structure would call for a careful re-interpretation of longitude-velocity $(\ell-v)$ diagrams \citep{Sofue+2025, Sofue+2025b}, which are foundational to our current models of gas flows and dynamics in the CMZ.

The existence of a nuclear bar in the MW would also have profound implications for our understanding of the growth history of the central super-massive black hole (SMBH), SgrA$^*$. In galaxies without recent major mergers such as the MW, SMBH growth is expected to proceed via secular processes. In single-barred systems, gas inflow driven by the primary bar typically stalls at a star-forming nuclear ring at radii of a few hundred parsecs, acting as a barrier that suppresses substantial accretion onto the SMBH \citep[e.g.,][]{Fanali+2015, Li+2017, Tress+2020}. By contrast, nuclear bars have been proposed to torque gas to smaller radii, channelling material toward the Galactic centre and potentially contributing to fuelling the central SMBH \citep{Shlosman+1989, Friedli+1993, Maciejewski+2002, Namekata+2009, Hopkins+2010, Li+2023}.

\subsection{Formation scenarios for a nuclear bar}

Numerical simulations show that in isolated systems, large-scale bars can form through dynamical instabilities in pre-existing discs \citep[see][and references therein]{Sellwood+2014}. The formation mechanism of nuclear bars is less well understood but is thought to be analogous, arising within a pre-existing nuclear stellar disc (NSD) \citep[see][for a review]{Schultheis+2025}. For instance, in \PaperI, the nuclear bar model evolved from the axisymmetric NSD model of \citet{Sormani+2022}, which was mildly unstable and spontaneously developed a nuclear bar. Other studies have similarly reproduced long-lived nuclear bars through such instabilities \citep[e.g.,][]{ Debattista+2007, Du+2015, Saha+2013, Wozniak+2015}, with external tidal perturbations potentially accelerating the formation process \citep{Semczuk+2024}. Observationally, nuclear bars are found at arbitrary orientations relative to the large-scale bar \citep{Buta+1993, Friedli+1993}, suggesting independent rotation. Simulations confirm this decoupling, also revealing pulsating amplitudes and pattern speeds as the nuclear bar rotates independently through the large-scale bar \citep{Debattista+2007, Li+2023}.

The relative timing of nuclear and large-scale bar formation also remains uncertain: simulations show nuclear bars can emerge either before \citep{Debattista+2007, Du+2015, Semczuk+2024} or after \citep{Wozniak+2015} their large-scale counterparts. There is however some observational evidence that stellar populations in nuclear bars are systematically younger than those of large-scale bars \citep{Lorenzo+2013, Lorenzo+2020}, favouring a formation sequence where the large-scale bar forms before the nuclear bar.

In the MW, observational evidence suggests the large-scale bar formed approximately $8\, \mathrm{Gyr}$ ago \citep[e.g.,][]{Bovy+2019, Haywood+2024, Sanders+2024}, though a much younger bar has also been proposed \citep[e.g.,][]{Cole+2002, Tepper+2021, Nepal+2024}. Within the NSD, \citet{Nogueras-Lara+2023} found significant differences in the stellar populations at different Galactocentric radii, revealing the presence of an age gradient, with intermediate-age populations $(2\mhyphen7\, \Gyr)$ more prominent in the outer NSD. This supports the scenario where the large-scale bar first funnelled gas inward to build the NSD, which could have subsequently become unstable and formed the nuclear bar through dynamical instabilities. A question which at present remains unclear is whether an age gradient would survive bar-formation and subsequent mixing from resonant interactions, especially given the short dynamical times near the Galactic centre. This may explain why \citet{Sanders+2024} found weak or no evidence for an age gradient when using Mira variables to trace the NSD's stellar age distribution.

There may also be the possibility that a nuclear bar existed in the past but has now dissolved. However, simulations suggest that nuclear bars dissolve when the central mass concentration exceeds roughly $0.1\%$ of the disc stellar mass \citep{Du+2017, Nakatsuno+2024}. The combined mass of SgrA$^*$ and the nuclear star cluster is reported to be below this threshold \citep{Nakatsuno+2024}, consistent with the potential survival of a nuclear bar in the MW.

\subsection{Nuclear bar parameters}

\begin{figure}
   \centering
    \includegraphics[width=.9\linewidth]{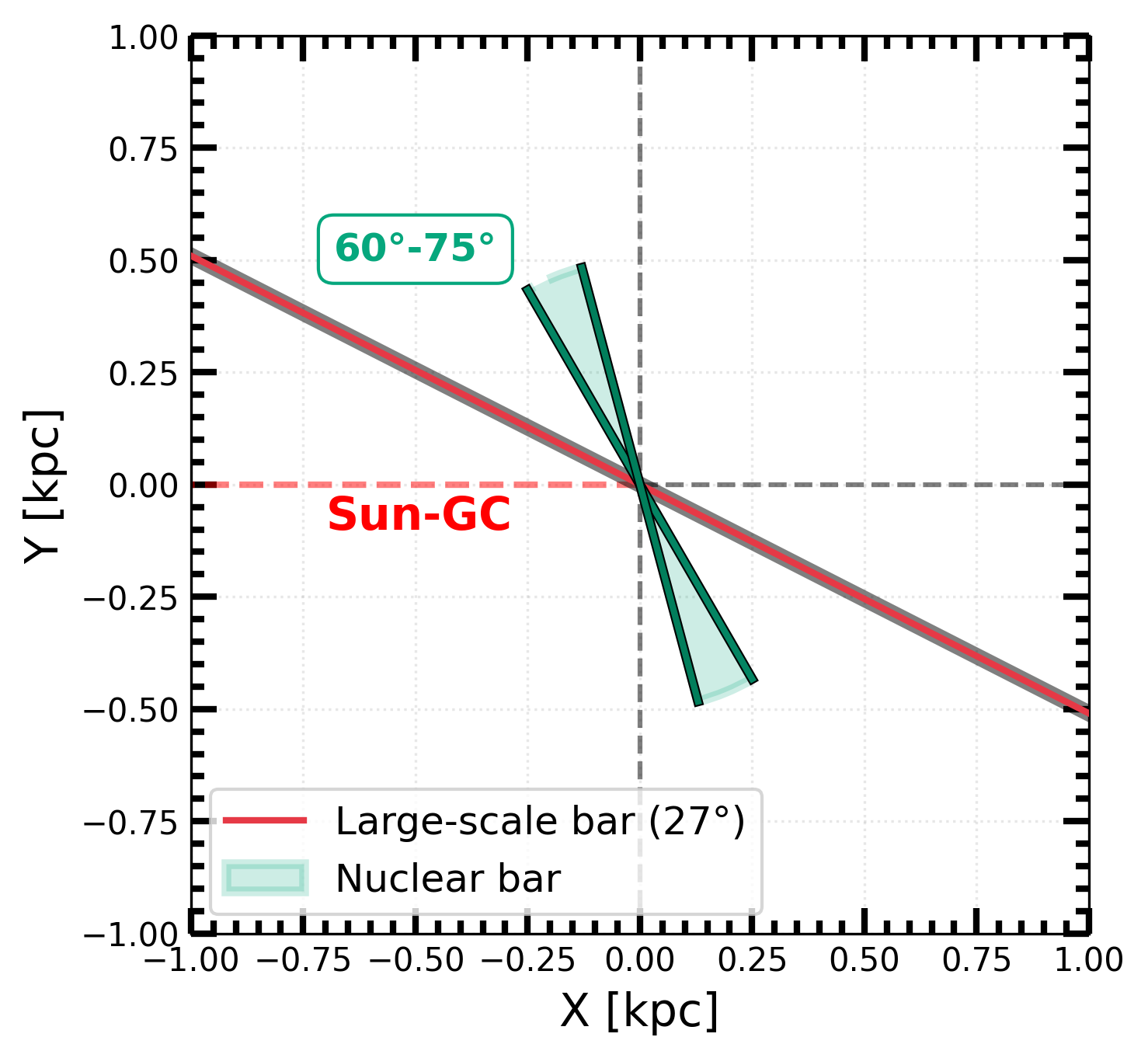}
    \caption{Schematic overview indicating the possible angles of the nuclear bar, based on the vertex deviation measurement obtained from sample D. The large-scale bar (red) is oriented at $27\degree$ to the Sun-GC line, while the inferred nuclear bar orientation (green lines) spans $60\degree\mhyphen75\degree$.}
    \label{fig:bar_angles}
\end{figure}

\textbf{Nuclear bar orientation:} In the presence of a bar, \lv\ depends on the angle which the bar makes with the Sun-GC line \citep[e.g.,][]{Simion+2021}. We compare the result for sample~D with the $l_v$--$\alpha$ relation from \PaperI\ (their Figure~12), which shows that $l_v$ varies smoothly with the nuclear bar orientation angle $\alpha$ and is a robust orientation diagnostic. A vertex deviation value of $l_v = -64.3^{+12.1}_{-12.2}\,^\circ$ roughly corresponds to a nuclear bar angle between $\approx 60^\circ$ and $75^\circ$. To illustrate this configuration, we plot in Fig.~\ref{fig:bar_angles} a schematic diagram with Solar position $(X,Y)=(-8.2,0)\kpc$. The large-scale bar (red line) is at an angle of $27\degree$ to the Sun-GC line \citep{Wegg+2015}. The two green lines represent the range of possible angles of the nuclear bar, as determined from the vertex deviation measurement of sample D. This orientation is consistent with the nuclear bar angle of $\alpha \approx 60\degree \mhyphen 75\degree$ reported by \citet{Rodriguez+2008} from gas dynamics modelling.

\noindent \textbf{Nuclear bar size:} In \PaperI, we showed that the \lv\ signal from the nuclear component dominates within $|\ell| < 1\degree$, though contamination from outer bar stars at larger longitudes prevents precise edge determination. Assuming a Galactic Center distance of $8.2\,\mathrm{kpc}$ and a viewing angle of $\alpha \approx 60-75\degree$, this yields a de-projected semi-major axis of $R=148-165\,\mathrm{pc}$, consistent with the MW's NSD size range of $100-230\,\mathrm{pc}$ \citep[e.g.,][]{Launhardt+2002,Gallego-cano+2020,Sormani+2022}.

\begin{figure}
   \centering
    \includegraphics[width=\linewidth]{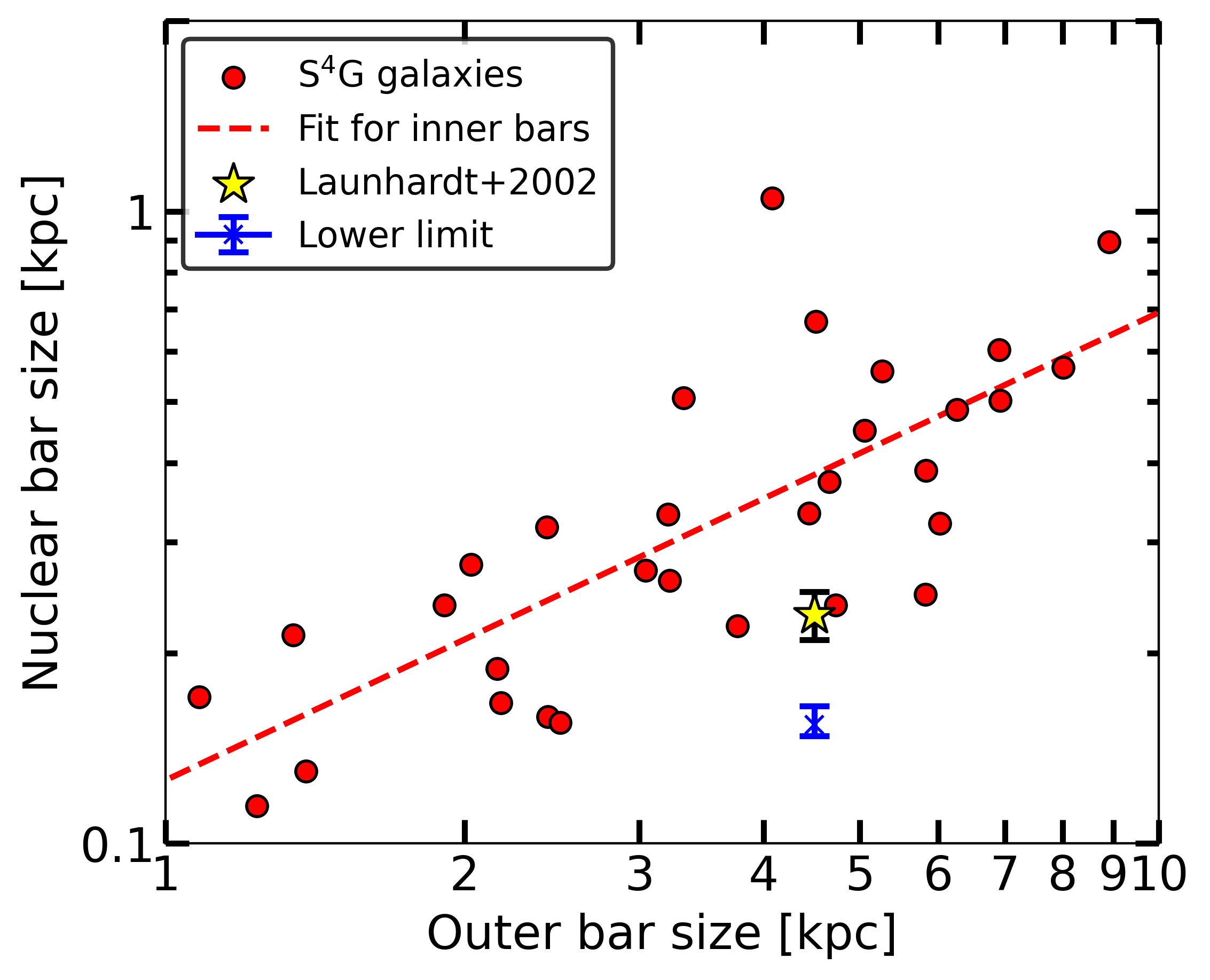}
    \caption{Figure reproduced from \citet{Erwin+2024} showing nuclear bar versus outer bar radius for external double-barred galaxies from the S$^4$G survey (red markers). The dashed line shows a power-law fit for the data. The blue marker represents the de-projected nuclear bar size ranging between $R=148\mhyphen165 \pc$ from this study. The yellow marker shows the NSD size reported by \citet{Launhardt+2002}. To estimate the outer bar size for the blue and yellow markers, we apply the ellipse fitting technique of \citet{Erwin+2004} to the made-to-measure model of the MW outer bar of \citet{Portail+2017}, obtaining a semi-major axis length of $4.5\kpc$.}
    \label{fig:radii}
\end{figure}

\citet{Erwin+2024} identified a strong correlation between inner and outer bar sizes in 31 double-barred S$^4$G galaxies (Fig.~\ref{fig:radii}, red markers and fit\footnote{https://github.com/perwin/db-nr\_paper/}). To obtain an estimate for the outer bar length, we applied the \citet{Erwin+2004} ellipse fitting method, defining bar length as the radius of maximum ellipticity, to the \citet{Portail+2017} made-to-measure bar model, yielding an outer bar semi-major axis of $4.5\,\mathrm{kpc}$  \citep[see][]{Wegg+2015}. Our nuclear bar measurement ($R=148-165\,\mathrm{pc}$, blue marker) lies $\sim2.6\sigma$ below the established trend. The yellow marker shows \citet{Launhardt+2002}'s NSD size, which sits at the lower end of the relation. Our value likely underestimates the true extent: while \PaperI\ models show the nuclear component's deviation signal dominates within $|\ell|<1\degree$, contamination from outer bar stars obscures the actual edge from our embedded perspective within the MW.

\noindent \textbf{Nuclear bar strength:} The strength of a bar is closely related to the anisotropy of the velocity ellipse, with stronger bars producing more elongated velocity distributions along their major axes (\PaperI, their Sec.~3.5). Our measurements yield conflicting constraints on the nuclear bar strength. Sample A, which encompasses the broader central region, shows low anisotropy ($\beta = 0.16^{+0.08}_{-0.05}$), suggesting a weak nuclear bar. In contrast, sample D, restricted to the innermost four fields, shows notably higher anisotropy ($\beta = 0.38^{+0.12}_{-0.07}$), indicative of a stronger nuclear bar. It is unclear whether this discrepancy reflects the statistical variations arising from smaller sample sizes, particularly Sample D, or the varying levels of contamination from the main bar across different fields. Disentangling these possibilities will require larger datasets with improved spatial coverage from upcoming surveys.

\subsection{Are we observing a nuclear bar?}

In summary, the key points which support the presence of a nuclear bar in the MW are:

\begin{enumerate}
    \item \textbf{Inconsistent with axisymmetry}: Both samples A and D exhibit strongly negative vertex deviations (\lv$=-54.8^{+13.1}_{-14.8} \,^\circ$ and $=-64.3^{+12.1}_{-12.2} \,^\circ$ , respectively) that are inconsistent with an axisymmetric structure, which would produce \lv$\approx 0\degree$ or $\pm 90\degree$.

    \item \textbf{Incompatible with an orthogonal elliptical NSD}: A slightly elliptical NSD with its major axis oriented orthogonally to the large-scale bar, as commonly observed in external galaxies \citep{Bittner+2020}, would produce a positive vertex deviation, contrary to our observations.


   \item \textbf{Distinct from the large-scale bar:} The samples defined in Sec.~\ref{sec:data} probe progressively cleaner subsets of the central NSD region. Sample~B (peripheral fields, bar-dominated) yields $l_v = -29.8^\circ$, whereas Samples~A and~D (central, $[\mathrm{Fe/H}] > -0.3$) yield $-54.8^\circ$ and $-64.3^\circ$, respectively, with a separation of $\approx 1.6\sigma$ and $\approx 2.6\sigma$ from Sample~B. The kinematic signal is moreover insensitive to the specific metallicity threshold used (Sec.~\ref{sec:contamination_data}, Fig.~\ref{fig:vdev_Fe}). The velocity-dispersion anisotropy also differs: Samples~A and~D exhibit maximum dispersion oriented along Galactic longitude, opposite to the line-of-sight–dominated dispersion characteristic of the main bar.

    \item \textbf{Robust against extinction-driven incompleteness:} A photometric incompleteness, driven by the VIRAC2 saturation limit at $K_s \approx 10$~mag, is detectable in the sample for $A_{K_s} \gtrsim 1.9$ (Sec.~\ref{sec:ext_vertex}, Fig.~\ref{fig:completeness}), but it does not appear to propagate into the kinematic measurement: $l_v$ and $\beta$ remain consistent within their bootstrap uncertainties as the upper $A_{K_s}$ cut is tightened down to $A_{K_s} \approx 2.0$ (Fig.~\ref{fig:vdev_ext}). Below this value the analysis is limited both by small sample sizes and by a target-selection bias affecting low-extinction fields (Sec.~\ref{sec:ext_vertex}), preventing a clean test in this regime. Furthermore, \PaperI\ shows that the longitude-asymmetric Galactic-centre extinction, when applied to an axisymmetric NSD model, shifts $l_v$ towards positive values, opposite in sign to our measurement. These complementary arguments together suggest that extinction is unlikely to be the cause of the observed signal.
  
\end{enumerate}

\noindent Nonetheless, we also stress that while the kinematic signatures reported in this work are suggestive of non-axisymmetry, we do not conclusively rule out an axisymmetric NSD. What is clear from our analysis is that samples A and D trace a structure kinematically distinct from the large-scale bar observed in sample B. However, several caveats should be considered. The relatively low anisotropy ($\beta = 0.16^{+0.08}_{-0.05}$ for sample A) is consistent with both a weak nuclear bar and an axisymmetric NSD (see \PaperI). While the measured vertex deviation is inconsistent with expectations for an axisymmetric NSD (which would yield $l_v \approx 0\degree$ or $\pm 90\degree$), it is possible that contamination from main bar stars is higher than anticipated, which could lead to a negative \lv. On the other hand, the notably higher anisotropy observed in sample D ($\beta = 0.38^{+0.12}_{-0.07}$) combined with a vertex deviation of $l_v = -64.3^{+12.1}_{-12.2}\,^\circ$ corroborates the case for streaming motions characteristic of a strong nuclear bar, but this sample is also the smallest ($N = 129$), and thus most susceptible to statistical fluctuations. We further note that the absolute spectroscopic completeness of the KMOS sample is low --- only $\sim$few percent of photometrically eligible stars in each field were actually targeted (Sec.~\ref{sec:ext_vertex}). While we assume this overall undersampling is approximately uniform within each field, our extinction-based robustness tests are not sensitive to potential biases introduced by the targeting strategy itself, and a definitive test will require independent kinematic surveys of the same region. In summary, while our findings favour the presence of a nuclear bar, a definitive confirmation will require the larger and more precise samples expected from upcoming surveys (Sec.~\ref{sec:future_prospects}). The kinematic signatures presented here, taken in combination with the diagnostics established in \PaperI, provide a compelling motivation for revisiting this analysis once such datasets become available.

\subsection{Future prospects}\label{sec:future_prospects}

Looking ahead, upcoming surveys targeting the Galactic nuclear region promise to provide a far more detailed and comprehensive view. In particular, the MOONS REDdened Milky WAY (REDWAY) survey \citep{Cirasuolo+2020,Gonzalez+2020} will provide high resolution spectra for $\sim13,000$ stars in $15$ independent fields within the region $|\ell|<1\degree$ and $|b|<0.5\degree$, enabling measurement of radial velocities, metallicities and $\alpha$-abundances. The \textit{Roman} Galactic Bulge Time Domain Survey \citep{Terry+2023} will provide high-precision photometric and astrometric measurements of $\sim3.3$ million stars over a $0.28$ deg$^2$ field centred on SgrA*, with proper motion precisions of $2.5-3.5\, \mu$as/yr. Lastly, JASMINE's Galactic Centre survey \citep{Kawata+2024} will cover a similar region $-0.7\degree<\ell<1.4\degree$ and $|b|<0.6\degree$, and is expected to provide accurate positions and proper motion measurements for about $1.2\times10^5$ stars up to $H_w < 14.5$ mag in the main survey area. A major step forward will come from the JWST/NIRCam Legacy Survey of the Galactic Center \citep{Schoedel+2023}, which will perform a large-area, deep multi-epoch NIRCam imaging survey of the inner $\sim$100 pc, delivering unprecedented stellar catalogues, individual star-by-star extinction measurements, and proper motion constraints across the NSD.

In parallel, Vasini et al.\ (in prep.) are developing a technique to constrain the three-dimensional distribution of gas and dust towards the Galactic centre. In addition to enabling future kinematic analyses to account for depth-dependent extinction and mitigate residual selection biases in low-extinction fields (Sec.~\ref{sec:ext_vertex}), this approach will also allow the detection of a possible nuclear bar directly from the gas distribution.

The large number of stars and the precision of measurements from these surveys will allow tight constraints on the level of non-axisymmetry in the NSD. Moreover, the availability of large datasets with metallicity, $\alpha$-abundances, and eventually stellar ages, will allow a more detailed exploration of how \lv, and $\beta$ vary across different stellar populations, yielding a more complete picture of the formation and evolution of this dynamically rich region of our Galaxy.

\begin{acknowledgements}
    We thank the anonymous referee for their valuable feedback, which has improved the quality of this manuscript. KF, MCS, XL, ZF, AV, and MD acknowledge financial support from the European Research Council under the ERC Starting Grant “GalFlow” (grant 101116226). MCS further acknowledges financial support from the Fondazione Cariplo under the grant ERC attrattivit\`{a} n. 2023-3014. We also acknowledge P. Erwin for providing public access to the data and code in Section 8.4, which improves our discussion of the results. We include a link to the data in the paper. FNL  and RS acknowledge financial support from the Severo Ochoa grant CEX2021-001131-S funded by MCIN/AEI/ 10.13039/501100011033 and from grants PID2022-136640NB-C21 and PID2024-162148NA-I00 funded by MCIN/AEI 10.13039/501100011033 and by the European Union. FNL acknowledges financial support from the Ramón y Cajal programme (RYC2023-044924-I) funded by MCIN/AEI/10.13039/501100011033 and FSE+. This study was based on data obtained from the ESO Science Archive Facility with DOI: https://doi.org/10.18727/archive/68 
    
\end{acknowledgements}

\bibliographystyle{aa}
\bibliography{refs2}

\end{document}